\documentclass{aa}  

\usepackage{hyperref}
\usepackage{stfloats}  % allows [!h] for double-column floats
\usepackage{graphicx}
\usepackage{orcidlink}
\usepackage{txfonts}
\begin{document} 

    \title{\texttt{ROLLIN'}: Rotating globular cluster simulations}
   \subtitle{I. The kinematic evolution of realistic direct N-body models}

   \author{P. Bianchini\inst{1}\orcidlink{0000-0002-0358-4502}
         \and
         A. L. Varri\inst{2,3}\orcidlink{0000-0002-6162-1594}
          \and
        A. Askar\inst{4}\orcidlink{0000-0001-9688-3458}
        \and
        A. Marklund\inst{1}\orcidlink{0009-0005-2779-8029}
        \and
         A. Mastrobuono-Battisti\inst{5,6,7}\orcidlink{0000-0002-2386-9142}
          }

   \institute{Universit\'e de Strasbourg, CNRS, Observatoire astronomique de Strasbourg, UMR 7550, F-67000 Strasbourg, France\\
              \email{paolo.bianchini@astro.unistra.fr}
      \and
              School of Mathematics and Maxwell Institute for Mathematical Sciences, University of Edinburgh, Kings Buildings, Edinburgh EH9 3FD, UK
        \and
                Institute for Astronomy, University of Edinburgh, Royal Observatory, Blackford Hill, Edinburgh EH9 3HJ, UK
      \and
            Nicolaus Copernicus Astronomical Center, Polish Academy of Sciences, ul. Bartycka 18, PL-00-716 Warsaw, Poland
        \and
         Dipartimento di Fisica e Astronomia “Galileo Galilei”, Università di Padova, Vicolo dell’Osservatorio 3, I-35122 Padova, Italy
         \and
         Dipartimento di Tecnica e Gestione dei Sistemi Industriali, Università di Padova, Stradella S. Nicola 3, I-36100 Vicenza, Italy
         \and
         Istituto Nazionale di Astrofisica – Osservatorio Astronomico di Padova, Vicolo dell’Osservatorio 5, Padova, I-35122, Italy 
             }

   \date{}

  \abstract
{
Internal rotation has recently emerged as a fundamental dynamical feature of globular clusters (GCs). Its presence can serve as a direct fossil record of GC formation and of their subsequent evolution within a galactic context, yet its origin and long-term evolution remain poorly understood.
}
{
In this paper, we systematically explore the evolution of rotating GCs over a Hubble time under the combined influence of two-body relaxation, external tidal field, and stellar evolution. Our goal is to establish a theoretical framework to interpret the growing wealth of 3D kinematic data and unveil primordial properties of GCs. 
}
{
We introduce the \texttt{ROLLIN’} simulations, a suite of 25 large N-body simulations of rotating GCs characterized by having a realistic number of stars from 250k to 1.5M and ran with the direct N-body code \texttt{NBODY6++GPU}. The simulations are evolved for 14~Gyr and are initialized with a wide range of rotation strengths, densities, and tidal fields. With present-day masses of $5\times10^4-5\times10^5$~M$_{\odot}$, the models cover the typical parameter space of low-density Milky Way GCs.
}
{
Our analysis highlights the pivotal role of rotation in shaping the early evolution of GCs. Rapidly rotating GCs experience earlier and more pronounced core collapse, efficiently segregating massive objects—including stellar remnants—in their centers within the first few hundred million years. In the long-term, internal rotation steadily declines, and after 12 Gyr, a correlation emerges between rotation strength and total GC mass, in agreement with recent observations. The primary driver of this evolution is mass loss, capturing both internal (stellar evolution, evaporation) and external processes (tidal stripping). The velocity anisotropy also evolves in response to mass loss: Clusters initially near isotropy develop radial anisotropy, peaking around 40\% mass loss, before progressing toward isotropy or tangential anisotropy at higher mass losses. Additionally, the GC orbital history plays a key role, as retrograde rotators retain angular momentum more effectively than prograde rotators, and their rotation profiles differ significantly already well within the Jacobi radius.
Finally, our simulations enabled us to quantify the long-term structural changes of GCs after 12 Gyr: (i) The surface density decreases by up to two orders of magnitude. (ii) The half-mass radius increases by a factor of three to five. (iii) The rotation strength decreases by a factor greater than five for clusters that have lost more than 50\% of their initial mass.
}
{The \texttt{ROLLIN’} simulations demonstrate that angular momentum, along with initial cluster density and mass, is one of the crucial aspects to understand the origin, evolution, and survival of GCs. These simulations therefore provide a valuable benchmark for interpreting GC observations both in the local and high-redshift Universe.
}


    \titlerunning{\texttt{ROLLIN'}: Rotating globular cluster simulations}
   \maketitle
%-------------------------
\section{Introduction}
%-------------------------
Globular clusters (GCs) are among the oldest and most compact stellar systems in the Universe, comprising roughly $10^6$ stars, and they are widely regarded as prominent fossils of the early cosmic epochs. Their ubiquitous presence across galaxies, together with the wealth of detailed observations available, makes GCs privileged probes of substructure formation at high redshift and of the subsequent tumultuous phases of galaxy assembly.

Modeling and interpreting the physical properties of these stellar systems remains a challenging task, as multiphysics processes—such as stellar and dynamical evolution—must be followed simultaneously over a Hubble time, from the small scales of individual stars to the large scales of the galactic environment. A particular difficulty arises from the fact that GCs are "collisional" systems in which frequent close encounters between stars play a major role in shaping their global properties, driving their evolution, and ultimately leading to their dissolution.

Direct summation, through direct integration of Newtonian gravitational forces, has proven to be a key method for investigating GC evolution (see review by \citealp{SpurzemKamlah2023}). For example, the \texttt{NBODY6} code (\citealp{Aarseth2003}) has enabled studies of the mechanisms driving GC mass loss \citep{BaumgardtMakino2003,BaumgardtSollima2017}, the formation of tidal tails (e.g., \citealp{Kupper2010,Webb2019}), the influence of time-varying external fields in galactic environments \citep{Bianchini2015,Miholics2016,Bianchini2017,Moreno-Hilario2024,Webb2024}, and processes such as binary star formation, the formation of stellar remnants, and the production of gravitational wave sources (e.g., \citealp{Sippel2013,ArcaSedda2024}).

Simulations of GCs with a realistic number of stars ($\approx10^6$) are now possible thanks to high-performance computing (HPC) facilities, which enable simultaneous modeling of dynamical evolution and stellar evolution in the presence of the external environment. Using the code \texttt{NBODY6++GPU} \citep{Wang2015}, one-to-one simulations of GCs with $N=10^5-10^6$ stars have been developed. Notable examples include the simulation presented in \citet{Sippel2013}, the simulation of M4 (\citealp{Heggie2014}), the suite of simulations \texttt{DRAGON-I} (\citealp{Wang2016}), \texttt{DRAGON-II} (\citealp{ArcaSedda2024}), and those from \citealt{Kamlah2022a}. These remarkable works have shown that the direct study of the long-term evolution of GCs remains challenging, with a realistic simulation still requiring computational times of more than $\sim1$ year even on parallel GPU architectures. However, alternative numerical methods relaxing the assumption of direct summation have proven to be extremely competitive in the task of simulating GCs with $10^6$ stars, particularly the code \texttt{PeTar} (\citealp{Wang2020}) and Monte Carlo method codes (the \texttt{MOCCA} code, see, e.g., \citealp{Hypki2013, Giersz2013}, and the \texttt{CMC} code, \citealp{Rodriguez2022}).

At the same time, observational facilities have advanced dramatically over the past $\sim10$ years, with high-quality data from both space- and ground-based telescopes providing precise kinematic measurements. These observations have revealed the unambiguous presence of internal rotation and velocity anisotropy in GCs, offering new constraints for theoretical models. Rotation was first measured in a few selected GCs thanks to line-of-sight observations (\citealp{Lane2011,Bellazzini2012,Bianchini2013,Kacharov2014,Fabricuius2014,Lardo2015,Cordero2017,Ferraro2018,Lanzoni2018}) and proper motion data (\citealp{vanLeeuwen2000,Anderson2003,Bellini2017,Haberle2024}). These types of measurements became feasible for an extended number of GCs thanks to large spectroscopic surveys (\citealp{Kamann2018,Szigeti2021,Martens2023,Petralia2024} and all-sky astrometric data from the \textit{Gaia} mission (\citealp{Bianchini2018b,Sollima2019,Vasiliev2019,VasilievBaumgardt2021}).
To date, the combination of these extensive datasets gives us access to the 3D rotational information for $\sim30$ GCs (e.g., \citealp{Leitinger2025} for a recent overview). The remarkable precision of these data has permitted measurements of rotation signals below the kilometer per second regime, of rotation differences between multiple stellar populations (\citealp{Cordoni2020,Martens2023,Dalessandro2024,Leitinger2025}), and between stars with different masses (\citealp{Scalco2023}). Velocity anisotropy has also been characterized on the plane of the sky for many GCs, showing that a variety of flavors are possible (from isotropy to mild radial and tangential anisotropy), using HST proper motions (\citealp{Watkins2015,Libralato2022}) and \textit{Gaia} data (e.g., \citealp{Jindal2019,VasilievBaumgardt2021,Cordoni2024}). The newly released high-precision astrometry and photometry from the Euclid mission seem particularly promising for carrying out further studies of this kind, particularly when combined with \textit{Gaia} data (see, e.g., \citealp{Libralato2024}).

On the modeling side, early efforts to understand rotation in GCs employed both direct N-body simulations and Fokker-Planck methods. Pioneering N-body studies explored the dynamical effects of rotation in self-consistent cluster models \citep{AkiyamaSugimoto1989,Ernst2007,Kim2008,Fiestas2010,Hong2013}.
At the same time, Fokker-Planck approaches \citep{EinselSpurzem1999,Kim2002,Fiestas2006,Tep2024} provided a complementary view of the long-term evolution, allowing researchers to track the influence of rotation on cluster structure and kinematics over extended timescales. Together, these early efforts set the stage for modern simulations that aim to capture the full complexity of rotating GCs.

Recent N-body simulations of rotating GCs have focused on the exploration of an increasing number of important physical ingredients, namely, the interplay between rotation and the tidal field (\citealp{Tiongco2018,Tiongco2022}), the effect of a stellar mass function (\citealp{Tiongco2021,Livernois2022}), the impact of stellar evolution (\citealp{Kamlah2022b}), the presence of multiple stellar populations (\citealp{Berczik2025,White2026}), and rotation in low-N clusters (\citealp{Bissekenov2025}). These simulations are limited because either they have a relatively low number of particles, $<10^5$, or they have only been evolved for a fraction of the Hubble time. Today, only one large rotating N-body model with $10^6$ stars, stellar evolution, and an external tidal field has been produced (\citealp{Wang2016}, \texttt{DRAGON-I}).

In this work, we explicitly aim to fill this gap by simulating realistic GCs with internal rotation for a Hubble time and for a wide set of initial conditions. The goal is to provide a description of present-day Milky Way (MW) GCs and to connect their current properties to their properties in the early Universe, which is now accessible via JWST high-redshift observations (\citealp{Vanzella2023,Adamo2024,Mowla2024}). In this paper, we present a suite of 25 rotating GC simulations with N=250k-1.5M stars run with the code \texttt{NBODY6++GPU}, which we name the  \texttt{ROLLIN'} simulations (ROtating globular cLusters Long-term evolutioN simulations). In particular, we focus on the characterization of their long-term kinematic evolution as an important step for interpreting and building a modeling benchmark for real GCs.

The paper is structured as follows. In Sect. \ref{sec:2} we describe the simulation setup, the parameter space explored, and the initial and final GC global properties. Section \ref{sec:3} is devoted to the early and long-term evolution of the clusters, with a particular emphasis on kinematic properties. In Sect. \ref{sec:4} we connect the present-day GC properties to their primordial properties, and in Sect. \ref{sec:5} we perform a comparison with available kinematic observations. Finally, in Sect. \ref{sec:6} we present our summary and conclusions.

%--------------------------------------------------------------------
\section{N-body simulations}
%--------------------------------------------------------------------
\label{sec:2}
We performed the simulations of rotating GCs using the code \texttt{NBODY6++GPU}, a GPU-accelerated direct N-body code \citep{Wang2015,Kamlah2022a} based on the family of N-body codes originally developed by Sverre Aarseth \citep{Aarseth1985,Spurzem1999,Aarseth1999, Aarseth2003, Nitadori2012}. The code allows the Newtonian equation of motion for an N-body system to be directly integrated and simultaneously includes a series of astrophysical ingredients, such as stellar and binary evolution and the interaction with an external tidal field.

\subsection{Initial conditions}
\label{sec:initial_conditions}
%-------------------------------------------------------------
%                   Two column Table 
%-------------------------------------------------------------
\begin{table*}
\caption{Initial conditions of the \texttt{ROLLIN'} simulations.}             
\label{table:1}      
\centering          
\begin{tabular}{l c c c c c c c c c l} 
\hline\hline       
Name & N & M & $r_{50\%}$ &$\Sigma_*$ & d & $r_{50\%}$/ $r_j$ &V$_{peak}$ & V$_{peak}/\sigma_0$ & log(t$_{rh}$/yr) & comments\\ 
 &   $10^5$ & $10^5$ $M_\odot$ & pc & 10$^3$ M$_\odot$\,pc$^{-2}$ & kpc & & km/s& & &\\ 

\hline                    
\texttt{1.5M-A-R4-10} & 15 & 8.58 & 4 & 8.5 &10 & 0.03 & 13.76 &1.22 & 9.30 & $-$\\ 
\texttt{500k-A-R2-10} &  5 & 2.87 & 2 &11.5 &10 & 0.02 & 11.26& 1.21 & 8.65& $-$\\ 
\texttt{500k-A-R4-10} &  5 & 2.88 & 4 & 2.9&10 & 0.04 & 7.96& 1.20 & 9.10 &$-$\\ 
\texttt{500k-A-R4-10-lC} &  5 & 2.88 & 4 & 2.9&10 & 0.04 &7.96 &1.20 & 9.10& \footnotesize{\texttt{level C} sev}\\ 
\texttt{500k-C-R4-10} &  5 & 2.86 & 4 & 2.8&10 & 0.04 & 5.53 &0.71 & 9.10 &$-$\\ 
\texttt{250k-A-R1-10} & 2.5 & 1.43 & 1 &22.8 & 10 & 0.01 &11.86 & 1.28 & 8.07 & $-$ \\
\texttt{250k-A-R2-25} &  2.5 & 1.44 & 2 &5.6& 25 & 0.01 &7.98  &1.22 & 8.53 &$-$\\ 
\texttt{250k-A-R2-25-vlk} &  2.5 & 1.44 & 2 &5.6& 25 & 0.01 & 7.98 &1.22 & 8.53&\footnotesize{$\sigma_{nk}=5$ km/s} \\ 
\texttt{250k-A-R2-10} &  2.5 & 1.44 & 2 & 5.6&10 & 0.03 & 7.98 &1.22 & 8.53& $-$\\ 
\texttt{250k-A-R2-5} & 2.5 & 1.44 & 2 & 5.6&5 & 0.05 & 8.31 &1.27 & 8.53&$-$ \\ 
\texttt{250k-A-R4-25} & 2.5 & 1.44 & 4 & 1.5&25 & 0.02 &5.63  &1.22 & 8.97& $-$\\ 
\texttt{250k-A-R4-25-imf50} & 2.5 & 1.37 & 4 & 1.4&25 & 0.02 & 5.50 &1.22 & 8.98& \footnotesize{$m_{max}=50$ M$_\odot$}\\ 
\texttt{250k-A-R4-25-lk} &  2.5 & 1.44 & 4 & 1.5&25 & 0.02 & 5.63 &1.22 & 8.97 &\footnotesize{$\sigma_{nk}=30$ km/s}\\ 
\texttt{250k-A-R4-25-vlk} &  2.5 & 1.44 & 4 &1.5 &25 & 0.02 & 5.63 &1.22 & 8.97& \footnotesize{$\sigma_{nk}=5$ km/s}\\ 
\texttt{250k-A-R4-25-retr} &  2.5 & 1.42 & 4 & 1.4&25 & 0.02 &$-$5.61  &$-$1.21 & 8.98 & \footnotesize{retrograde}\\ 
\texttt{250k-A-R4-10} &  2.5 & 1.44 & 4 & 1.5&10 & 0.05 & 5.63 &1.22 & 8.97 &$-$\\ 
\texttt{250k-A-R4-10-retr} &  2.5 & 1.42 & 4 & 1.4&10 & 0.05 & $-$5.61 &$-$1.21 & 8.98 &\footnotesize{retrograde}\\ 
\texttt{250k-B-R4-25} &  2.5 & 1.41 & 4 & 1.4&25 & 0.02 & 4.92 &0.99 & 8.98&$-$ \\ 
\texttt{250k-B-R4-25-lk} &  2.5 & 1.41 & 4 & 1.4&25 & 0.02 & 4.92 &0.99 & 8.98 &\footnotesize{$\sigma_{nk}=30$ km/s}\\ 
\texttt{250k-C-R2-10} &  2.5 & 1.43 & 2 & 5.6&10 & 0.03 & 5.52 &0.70 & 8.53&$-$ \\ 
\texttt{250k-C-R4-25}&  2.5 & 1.43 & 4 & 1.4&25 & 0.02 & 3.90 &0.71 & 8.97&$-$ \\ 
\texttt{250k-C-R4-25-lk} &  2.5 & 1.43 & 4 &1.4 &25 & 0.02 & 3.90 &0.71 & 8.97 &\footnotesize{$\sigma_{nk}=30$ km/s}\\ 
\texttt{250k-C-R4-10} & 2.5 & 1.43 & 4 &1.4 &10 & 0.05 & 3.90 &0.71 & 8.97 &$-$\\ 
\texttt{250k-W6-R4-25} &  2.5 & 1.42 & 4 &1.4 &25 & 0.02 & 2.36 &0.44 & 8.98 &$-$\\ 
\texttt{250k-W6-R4-25-retr} & 2.5 & 1.43 & 4 &1.4 &25 & 0.02 & $-$2.31  &$-$0.42 & 8.98 &\footnotesize{retrograde}\\ 

\hline                  
\end{tabular}
\tablefoot{Initial conditions of our set of simulations: the name of the simulation (see naming convention in Sect. \ref{sec:2}), the number of stars N, the total mass M, the half-mass radius $r_{50\%}$, the mass surface density within the half-mass radius $\Sigma_*$, the distance d to the central point-mass galaxy, the filling factor defined as the ratio between the half-mass radius and the Jacobi radius $r_j$, the peak of the rotation profile $V_{peak}$, the $V_{peak}/\sigma_0$ ratio with $\sigma_0$ the central velocity dispersion, and the log of the half-mass radius relaxation time $t_{rh}$. The last column indicates additional information on the simulations (as described in Sect. \ref{sec:2}), in particular nonstandard values of natal kicks $\sigma_{nk}$, \texttt{level C} stellar evolution, maximum value of the stellar mass function $m_{max}$ and retrograde rotation. 
}
\end{table*}
%

%-------------------------------------- Two column figure (place early!)
   \begin{figure*}
   \centering
   \includegraphics[width=1\textwidth]{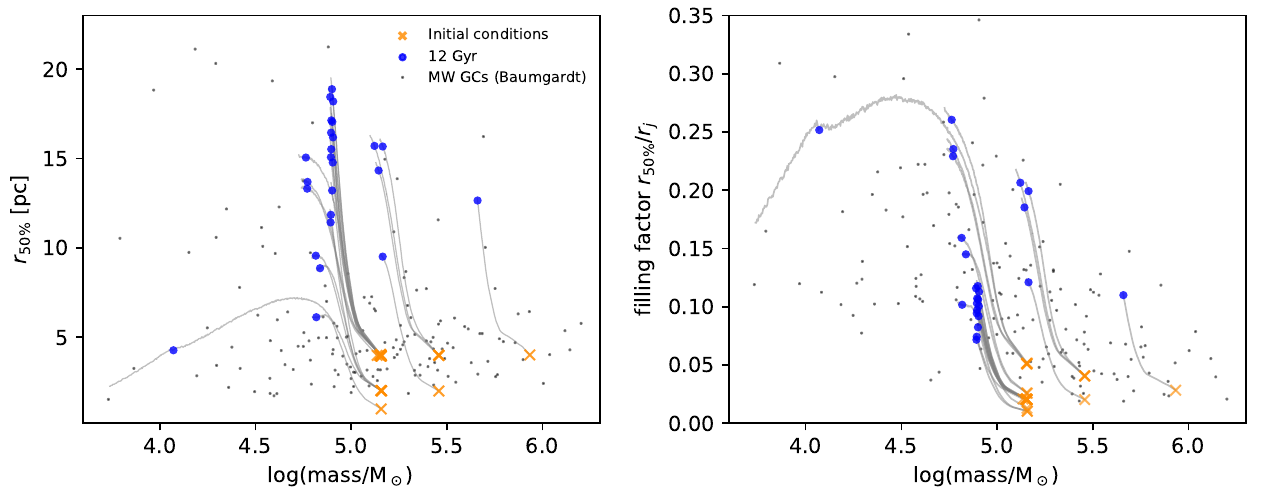}
   \caption{Time evolution of the mass; half-light radius, $r_{50\%}$; and filling factor, $r_{50\%}/r_j$, of our set of simulations. Orange crosses are the initial conditions, blue dots are the snapshots at 12 Gyr, and the gray lines indicate the time evolution. The comparison with MW GCs (gray dots, Baumgardt GCs Database) highlights that our simulated GCs cover the current parameter space of real GCs in the low-density regime.}
              \label{fig:ic}%
    \end{figure*}
%--------------------------------------------%

%-------------------------------------------------------------
%                                             Two column Table 
%-------------------------------------------------------------
\begin{table*}
\caption{Properties of the \texttt{ROLLIN'} simulations at 12 Gyr.}             
\label{table:2}      
\centering          
\begin{tabular}{l c c c c c c c c c l}      
\hline\hline       
                     
Name & N & M & $\Delta$M/M& $r_{50\%}$ & $\Sigma_*$ & $r_{50\%}$/ $r_j$ & V$_{peak}$ & V$_{peak}/\sigma_0$ & $\beta_{<50\%}$ & log(t$_{rh}$/yr) \\ 
 &  $10^5$ & $10^5$ $M_\odot$ & & pc & 10$^3$ M$_\odot$\,pc$^{-2}$& & km/s & & &\\ 

\hline                    
\texttt{1.5M-A-R4-10} & 13.59 & 4.58 & 0.47 & 12.65 & 0.45 &0.11 & 2.77 & 0.40 & 0.18 & 10.15 \\
\texttt{500k-A-R2-10} & 4.31 & 1.46 & 0.49 & 9.51 & 0.26 &0.12 & 1.49 & 0.36 & 0.15 & 9.75 \\
\texttt{500k-A-R4-10} & 4.04 & 1.39 & 0.52 & 14.33 &0.11 &0.19 & 1.09 & 0.30 & 0.15 & 10.00 \\
\texttt{500k-A-R4-10-lC} & 3.79 & 1.32 & 0.54 & 15.71 & 0.09&0.21 & 0.96 & 0.28 & 0.15 & 10.05 \\
\texttt{500k-C-R4-10} & 4.27 & 1.46 & 0.49 & 15.67 & 0.09&0.20 & 0.79 & 0.22 & 0.13 & 10.07 \\
\texttt{250k-A-R1-10} & 1.84 & 0.66 & 0.54 & 6.12 & 0.28 & 0.10 & 1.19 & 0.36 & 0.06 & 9.30 \\
\texttt{250k-A-R2-25} & 2.35 & 0.78 & 0.46 & 11.43 & 0.09&0.07 & 1.19 & 0.43 & 0.18 & 9.77 \\
\texttt{250k-A-R2-25-vlk} & 2.33 & 0.80 & 0.45 & 13.21 & 0.07&0.08 & 1.05 & 0.39 & 0.21 & 9.85 \\
\texttt{250k-A-R2-10} & 1.85 & 0.65 & 0.55 & 9.57 &0.11 &0.16 & 0.75 & 0.27 & 0.12 & 9.60 \\
\texttt{250k-A-R2-5} & 0.20 & 0.12 & 0.92 & 4.28 &0.10 &0.10 & $-$0.09 & 0.06 & $-$0.06 & 8.57 \\
\texttt{250k-A-R4-25} & 2.35 & 0.79 & 0.45 & 15.52 &0.05 &0.10 & 1.10 & 0.45 & 0.18 & 9.97 \\
\texttt{250k-A-R4-25-imf50} & 2.37 & 0.78 & 0.43 & 11.85 & 0.04&0.07 & 1.33 & 0.49 & 0.15 & 9.79 \\
\texttt{250k-A-R4-25-lk} & 2.34 & 0.79 & 0.45 & 17.13 &0.04& 0.11 & 1.02 & 0.43 & 0.20 & 10.03 \\
\texttt{250k-A-R4-25-vlk} & 2.34 & 0.81 & 0.44 & 18.19 & 0.04&0.11 & 0.97 & 0.40 & 0.21 & 10.06 \\
\texttt{250k-A-R4-25-retr} & 2.34 & 0.78 & 0.45 & 15.07 & 0.05&0.09 & $-$1.31 & 0.53 & 0.17 & 9.94 \\
\texttt{250k-A-R4-10} & 1.64 & 0.59 & 0.59 & 13.31 & 0.05&0.23 & 0.46 & 0.19 & 0.11 & 9.78 \\
\texttt{250k-A-R4-10-retr} & 1.64 & 0.59 & 0.58 & 13.69 &0.10 &0.24 & $-$0.92 & 0.41 & $-$0.02 & 9.80 \\
\texttt{250k-B-R4-25} & 2.41 & 0.80 & 0.43 & 14.78 & 0.12&0.05 & 1.12 & 0.45 & 0.17 & 9.94 \\
\texttt{250k-B-R4-25-lk} & 2.39 & 0.80 & 0.43 & 16.18 &0.05 &0.10 & 1.04 & 0.42 & 0.18 & 9.99 \\
\texttt{250k-C-R2-10} & 1.98 & 0.69 & 0.52 & 8.87 &0.14 &0.14 & 0.56 & 0.19 & 0.07 & 9.56 \\
\texttt{250k-C-R4-25} & 2.39 & 0.80 & 0.44 & 17.05 & 0.04&0.11 & 0.78 & 0.32 & 0.17 & 10.03 \\
\texttt{250k-C-R4-25-lk} & 2.36 & 0.79 & 0.44 & 18.88 &0.04 &0.12 & 0.69 & 0.29 & 0.18 & 10.09 \\
\texttt{250k-C-R4-10} & 1.62 & 0.58 & 0.59 & 15.05 & 0.04&0.26 & 0.22 & 0.10 & 0.09 & 9.86 \\
\texttt{250k-W6-R4-25} & 2.36 & 0.79 & 0.45 & 16.45 & 0.05&0.10 & 0.24 & 0.10 & 0.16 & 10.00 \\
\texttt{250k-W6-R4-25-retr} & 2.32 & 0.78 & 0.46 & 18.44 & 0.04&0.12 & $-$0.50 & 0.21 & 0.18 & 10.07 \\

\hline                  
\end{tabular}
\tablefoot{Properties of the simulations at 12 Gyr, as in Table \ref{table:1}. Additionally, we report the values of the mass loss $\Delta M/M$ and the anisotropy parameter $\beta$ within the half-mass radius.}
\end{table*}
As initial conditions, we adopted the family of distribution function–based models developed by \cite{VarriBertin2012} and constructed to describe quasi-relaxed stellar systems with realistic rotation. These models feature axisymmetry, differential rotation, and a velocity anisotropy profile that is isotropic at the center, radially anisotropic in the intermediate regions, and mildly tangentially anisotropic in the outskirts. 

The models are expressed as functions of the integral of motion $I=I(E,J_z)$, with $E$ and $J_z$ as the energy and z-component of the angular momentum, defined as
\begin{equation}
\label{eq:1}
I=E-\frac{\omega J_z}{1+bJ_z^{2c}},
\end{equation}
The models are characterized by four dimensionless parameters: 
\begin{itemize}
    \item the concentration parameter, $W_0$, defined as the central depth of the dimensionless potential well;
    \item the rotation strength parameter, $\hat{\omega}$, controlling the amount of rotation present in the cluster;
  \item the parameters $b$ and $c$ defining the shape of the rotation profile.
\end{itemize}
For a full description of each parameter, we refer to \citealp{VarriBertin2012}.\footnote{We note that in other studies (e.g., \citealp{Bianchini2013}) the rotation strength parameter for this family of models is indicated as $\chi$, defined as $\chi=\omega^2/(4\pi G\rho_0)$. The parameters $\omega$ and $\chi$ are connected to our dimensionless rotation parameter $\hat{\omega}$ through this relation: $\hat{\omega}=3\chi^{1/2}$.}

To construct initial conditions representative of real GCs, we based our choice of the four dimensionless parameters on the model–observation comparisons presented in \citet{Bianchini2013} and \citet{Kacharov2014}. In these works, the typical structural parameters derived from the best fit models are $W_0 = 7$, $b = 0.01$, and $c = 1$.\footnote{Note that a similar choice was made in \citep{Tiongco2016b}.} While these parameters are kept fixed, we further explore three different values of the rotation parameter, $\hat{\omega} = 0.3, 0.2, 0.1$, which correspond to progressively decreasing rotation strengths. Models with higher $\hat{\omega}$ are characterized by a stronger rotational support, quantifiable as a higher $V/\sigma$ parameter, as indicated in Tab.~\ref{table:1}.

The initial conditions are characterized by three values of initial number of stars, $N = (2.5,5,15) \times 10^5$, and two values for the initial half-mass radius $r_{50\%}=1$, 2 and 4 pc.\footnote{To our best knowledge, the simulation with 1.5M stars is the largest direct N-body model developed so far.} The initial mass function (IMF) is assigned following a \cite{Kroupa2001} mass function with minimum and maximum stellar mass of 0.08 and 100 $M_\odot$ and with no primordial binaries. 
Previous simulations of rotating GCs often consider an initial mass function truncated at lower stellar masses (e.g., \citealp{Livernois2022} use a Kroupa mass function with a mass range of 0.1-1 M$_\odot$). To allow for an exploration of the effects due to the presence or absence of high mass stars, we set an additional simulation with maximum stellar mass of 50 $M_\odot$.
The stellar metallicity is set to [Fe/H]=$-1.3$ (i.e., Z=0.001). All clusters are evolved in circular orbits with distances of $d=5$, 10, 25, 50 kpc, around an idealized point-mass galaxy with mass $M=10^{11}$ $M_\odot$. The rotation axes of the GCs are initialized along the z-axis, which is perpendicular to the orbital plane x-y around the point-mass galaxy. The majority of simulations are set with prograde internal rotation with respect to the orbital motion of the GC in the galaxy (i.e., the internal angular momentum vector is aligned with the orbital angular momentum vector), while three simulations are set with retrograde rotation (i.e., rotation axis along the negative direction of the z-axis). These orbital properties set all clusters in the tidal underfilling regime, with filling factors (i.e., the ratio between the half-mass ratio and the Jacobi radius, $r_j$\footnote{In case of circular orbit around a point mass galaxy the Jacobi radius is defined as $r_j=d\,(M_{GC}/3M_{gal})^{1/3}$; see, e.g., \citep{Renaud2011}.}) of $r_{50\%}/r_j\sim0.01-0.05$. 

The implementation of stellar evolution for single and binary stars follows the prescriptions defined in \cite{Wang2016}, based on the original SSE and BSE prescriptions of \citep{Hurley2000,Hurley2002}, commonly referred in \texttt{NBODY6++GPU} as \texttt{level A} (we refer to \citealp{Kamlah2022a} for details on the implementation of stellar evolution). In this context, for most of the simulations, the natal kick experienced by neutron stars at formation is set as a Maxwellian distribution with a dispersion of $\sigma_{nk}=265$~km/s (see \citealp{Hobbs2005}; but also see \citealp{Disberg2025}). In order to explore different stellar remnant retentions, we employ for three simulations a natal kick dispersion of $\sigma_{nk}=30$ km/s (low natal kick), and for two simulations $\sigma_{nk}=5$ km/s (very low natal kick). For all simulations the natal kick of black holes is modeled including a fallback mechanism (see \citealp{Belczynski2002}). To allow for a comparison with the most updated version of the stellar evolution code, we ran one further simulation with 500k stars using the \texttt{level C} stellar evolution implementation.\footnote{Note that in the version of the code used for this paper, the kick mechanism in \texttt{level C} is set to "collapse-asymmetry-driven," that is the parameter \texttt{KMECH}=3. See \cite{Kamlah2022a} for details.} The effects of the natal kick and stellar evolution prescriptions, and in particular their connection to the number of stellar remnants retained by the GCs, will be examined in a follow-up paper.

Finally, we run two control simulations with a different rotating model as initial condition. Instead of using the distribution function of \cite{VarriBertin2012}, we set a spherical, isotropic and nonrotating \cite{Wilson1975} model with $W_0=6$, in which solid body rotation is introduced using the "Lynden-Bell demon" \citep{Lynden-Bell1962}. This method corresponds to imposing positive values to the $v_\phi$ velocity component for a fraction of stars that otherwise had negative values (see, e.g., recent applications \citealp{Rozier2019,Breen2021}). In our case, we randomly select 50\% of the stars in our sampled distribution function and impose the absolute value of $v_\phi$ and consider one prograde and one retrograde initial condition.

The naming convention used to identify the simulations, as reported in Table \ref{table:1}, explicitly describes the initial conditions. The format is the following: number of stars N, model type (indicated as \texttt{-A}, \texttt{-B}, or \texttt{-C} for the \citealp{VarriBertin2012} models with rotation strengths of $\hat{\omega}=0.3, 0.2, 0.1$ respectively, and \texttt{-W6} for the rotating Wilson models), half-mass radius in parsec, size of the circular orbit in kpc, plus complementary information if a property differs from the standard one. This additional information includes the use of \texttt{level C} stellar evolution (indicated as \texttt{-lC}), low or very low natal kicks of stellar remnants (indicated as \texttt{-lk} or \texttt{-vlk}), truncation of the IMF to 50 $M_\odot$ (indicated as \texttt{-imf50}), and retrograde internal rotation with respect to the galactic orbital motion (\texttt{-retr}). As an example, \texttt{1.5M-A-R4-10} designates the simulation with 1.5M stars, rotating \cite{VarriBertin2012} model with $\hat{\omega}=0.3$, half-mass radius $r_{50\%}=4$ pc and circular orbit with $d=10$ kpc.

The suite of simulations consists of a total of 25 GC models evolved up to 14 Gyr, with the exception of the simulation \texttt{1.5M-A-R4-10} which is run until 12 Gyr. The simulations were run on the supercomputer Jean-Zay (GENCI–IDRIS) on NVIDIA V100 GPUs over a $\sim3$ yr period, under the time allocations Grand Challenge-101470 (08/2020), A10-A0100412451 (05/2021), and A13-A0130412451 (11/2022). The entire set of simulations was run for a total of $\sim350\,000$ GPU hours. A typical simulation with 250k stars took approximately 20-50 days with 4-16 GPUs, while the largest 1.5M star simulation took $\sim400$ days with 12-48 GPUs. The total CO$_2$ equivalent consumption\footnote{The carbon footprint is calculated using the methodology developed by the Labos 1point5 collective, \url{https://labos1point5.org/les-rapports/estimation-empreinte-calcul}.} for our simulations is $\sim9$ tons of CO$_{2,eq}$.

%-------------------------------------- Two column figure (place early!)
   \begin{figure*}
   \centering
   \includegraphics[width=1\textwidth]{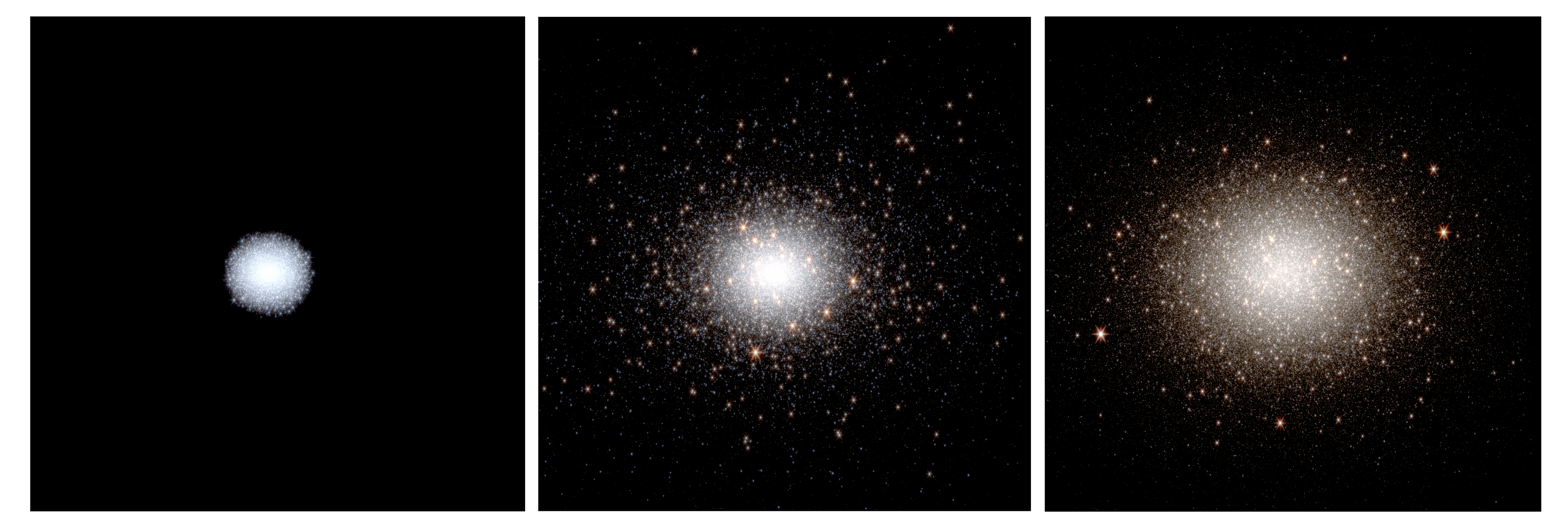}
   \caption{Visualization of the \texttt{1.5M-A-R4-10} model as if it were observed by the JWST NIRcam with the F070W, F115W, and F356W filters. The three images correspond to snapshots at 0, 1, and 12 Gyr observed at a distance of d=200 kpc and with an inclination angle between the line of sight and the rotation axis of $i=45^\circ$ and a field of view of 132x132 arcsec$^2$. A video illustrating the time evolution of this simulation is \href{https://www.youtube.com/watch?v=o_C2nwJq560}{available at this link}.}
              \label{Fig:image1}

   \centering
   \includegraphics[width=1\textwidth]{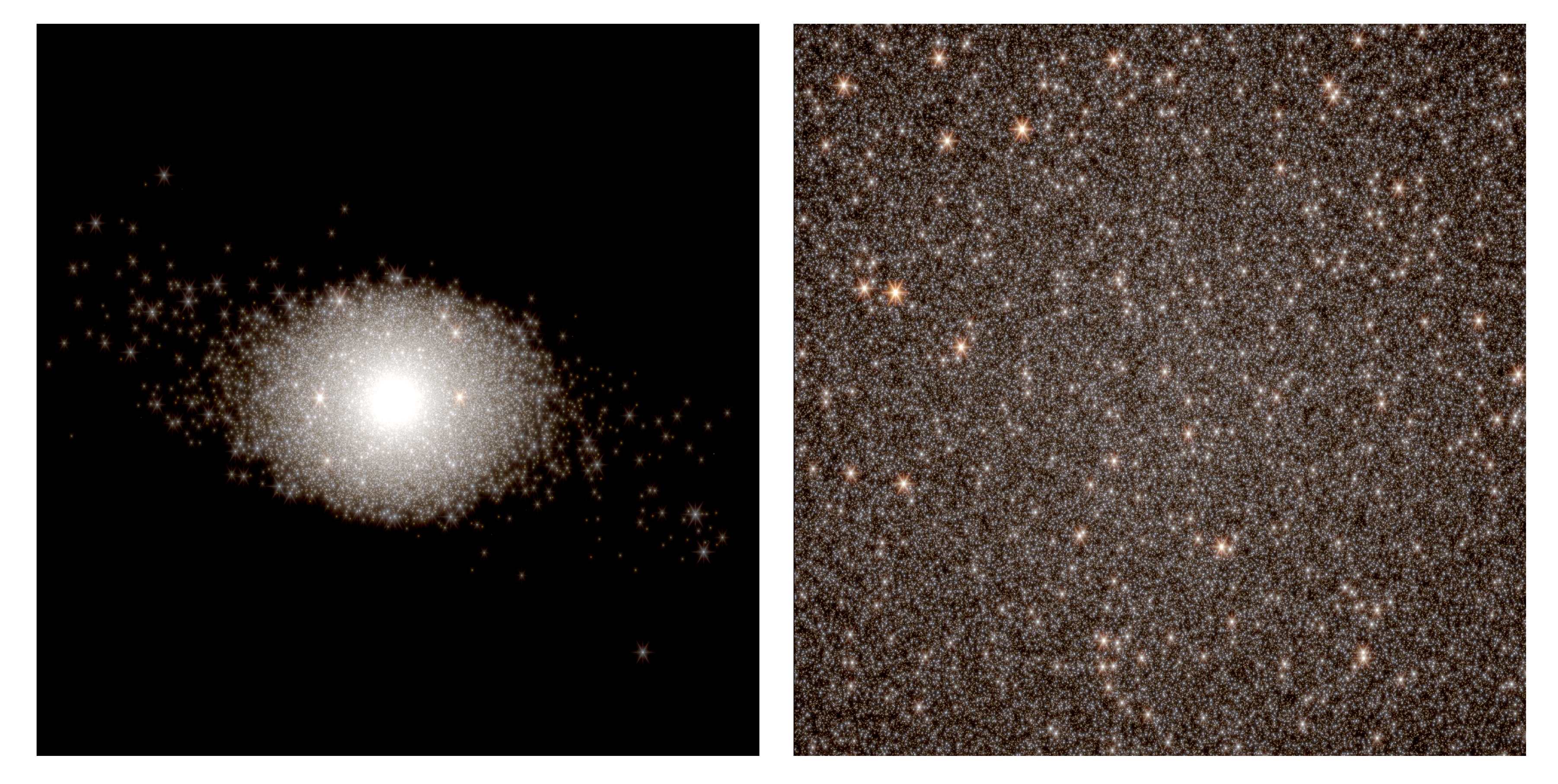}
   \caption{Visualization of the \texttt{1.5M-A-R4-10} model at 12 Gyr as if it were observed by the JWST NIRcam (with the F070W, F115W, and F356W filters) with a field of view of 132x132 arcsec$^2$ at two different distances in an extragalactic and galactic context (d=800 kpc in the left image, and d=20 kpc in the right image). The GC is observed face on, along the direction of the rotation axis, which is perpendicular to the orbital plane.}
              \label{Fig:image2}
    \end{figure*}
%--------------------------------------------%

%-------------------------------------- Two column figure (place early!)
   \begin{figure*}
   \centering
   \includegraphics[width=1\textwidth]{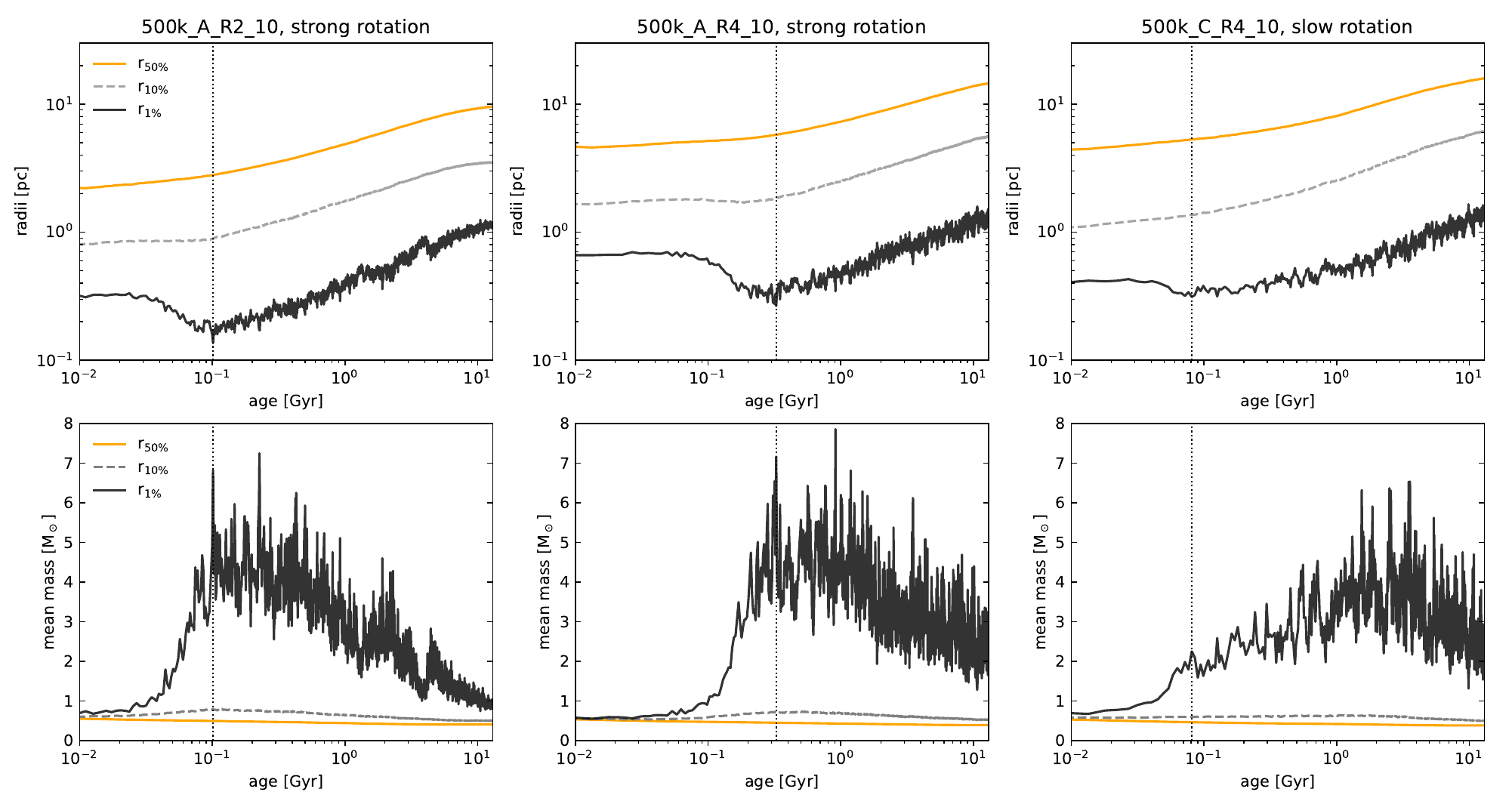}
   \caption{\textit{Top row:} Evolution of 50\%, 10\%, and 1\% Lagrangian radii (orange, gray, and black lines) for simulations \texttt{500k-A-R2-10}, \texttt{500k-A-R4-10}, and \texttt{500k-C-R4-10}. The three simulations are representative of different densities and rotation strengths. The vertical lines indicate the time of core collapse. Higher density GCs show an earlier and deeper collapse, whereas low rotating GCs show a weaker signature of collapse. \textit{Bottom row:} Evolution of the mean mass within the same Lagrangian radii of the top panel. The 1\% Lagrangian radii mainly contain massive objects, corresponding to BHs (or their progenitors). High-density and strong rotating GCs mass segregate more efficiently, in timescales of a few hundred million years.}
              \label{core_collapse}%
    \end{figure*}
%

%--------------------------------------------%

\subsection{General properties}
\label{sec:gen_propr}

The goal of this work is to investigate the long-term evolution of the kinematic and structural properties of rotating GCs, and their interplay with tidal effects and initial GCs' properties. Our grid of simulations was specifically selected to explore these effects and these dependencies. Figure \ref{fig:ic} shows the evolution of the mass, half-mass radius, and filling factor of the \texttt{ROLLIN'} simulations compared to the observed values of MW GCs (from the Baumgardt Galactic GCs Database\footnote{These parameters were first derived in \cite{BaumgardtHilker2018} and updated in \url{https://people.smp.uq.edu.au/HolgerBaumgardt/globular/}}). The snapshots at 12 Gyr of our simulations cover the parameter space of real MW GCs in the low-density regime. In essence, they allowed us to explore the following: 
\begin{itemize}
    \item a variety of initial mass surface densities ranging from $\Sigma_*\approx$~$10^3$ to $\Sigma_*\approx10^4$ $M_\odot / pc^2$;
    \item different strengths of the tidal field, including the effects of prograde versus retrograde cluster rotation with respect to their orbits in the galactic potential;
    \item different stellar evolution prescriptions (\texttt{level A} and \texttt{level~C}) and the effects of different natal kicks.
\end{itemize}
In Tables \ref{table:1} and \ref{table:2}, we report the summary of the initial and 12~Gyr properties, respectively. In particular, we include the estimate of internal rotation quantified by the peak of the rotation profile $V_{peak}$, the $V_{peak}/\sigma_0$ parameter with $\sigma_0$ the central value of the 1D velocity dispersion, and the $\beta$ anisotropy parameter defined as $\beta=1-(\sigma^2_\theta+\sigma^2_\phi)/2\sigma^2_r$, with ($r,\theta,\phi$) representing the components in spherical coordinates of the velocity dispersion. Note that, not reported in Table \ref{table:1}, the initial values of $\beta$ within the half-light radius for all the simulations are $\beta\approx0$, except for \texttt{250k-W6-R4-25} and \texttt{250k-W6-R4-25-retr} which have $\beta\approx0.08$. The relaxation time at the half-mass radius $t_{rh}$ is calculated using eq. 2 in \citet{Bianchini2016}.

In Figs. \ref{Fig:image1} and \ref{Fig:image2}, we show a visualization of the \texttt{1.5M-A-R4-10} simulation as if it were observed with a JWST NIRcam setup. The color images were produced with three channels using the F070W, F115W, and F356W photometric filters, and using a filter-dependent PSF produced with the Python package \texttt{WebbPSF} \citep{perrin2012}.\footnote{\url{https://webbpsf.readthedocs.io/en/latest/}} The stellar magnitude of each star was calculated from the stellar parameters (luminosity, mass, stellar radius; contained in the outputs of the simulations) using the software \texttt{FSPS} \citep{Conroy2009,Conroy2010}\footnote{\url{https://github.com/cconroy20/fsps}} and employing BaSel spectra library and Padova isochrones library. Finally, the field of view was defined to have 132x132 arcsec$^2$ size, with a pixel scale of 0.062 arcsec. Figure \ref{Fig:image1} displays three snapshots, at 0, 1, and 12 Gyr, observed at a distance of 200 kpc and with an inclination angle between the line-of-sight at the rotation axis of i=45$^\circ$. These panels show the effect of dynamical evolution (e.g., expansion of the clusters) and of stellar evolution (e.g., change of color of the stellar population). Figure \ref{Fig:image2} shows instead the 12 Gyr snapshots observed face on (along the rotation axis, perpendicular to the orbital plane of the cluster) at different distances, underlying the amount of star-by-star details available in the dense central region as well as the large-scale effects of the tidal field (e.g., formation of tidal features). A video illustrating the time evolution of this simulation is also available \href{https://www.youtube.com/watch?v=o_C2nwJq560}{at this link}.

%-------------------------------------- Two column figure (place early!)
   \begin{figure*}
   \centering
   \includegraphics[width=1\textwidth]{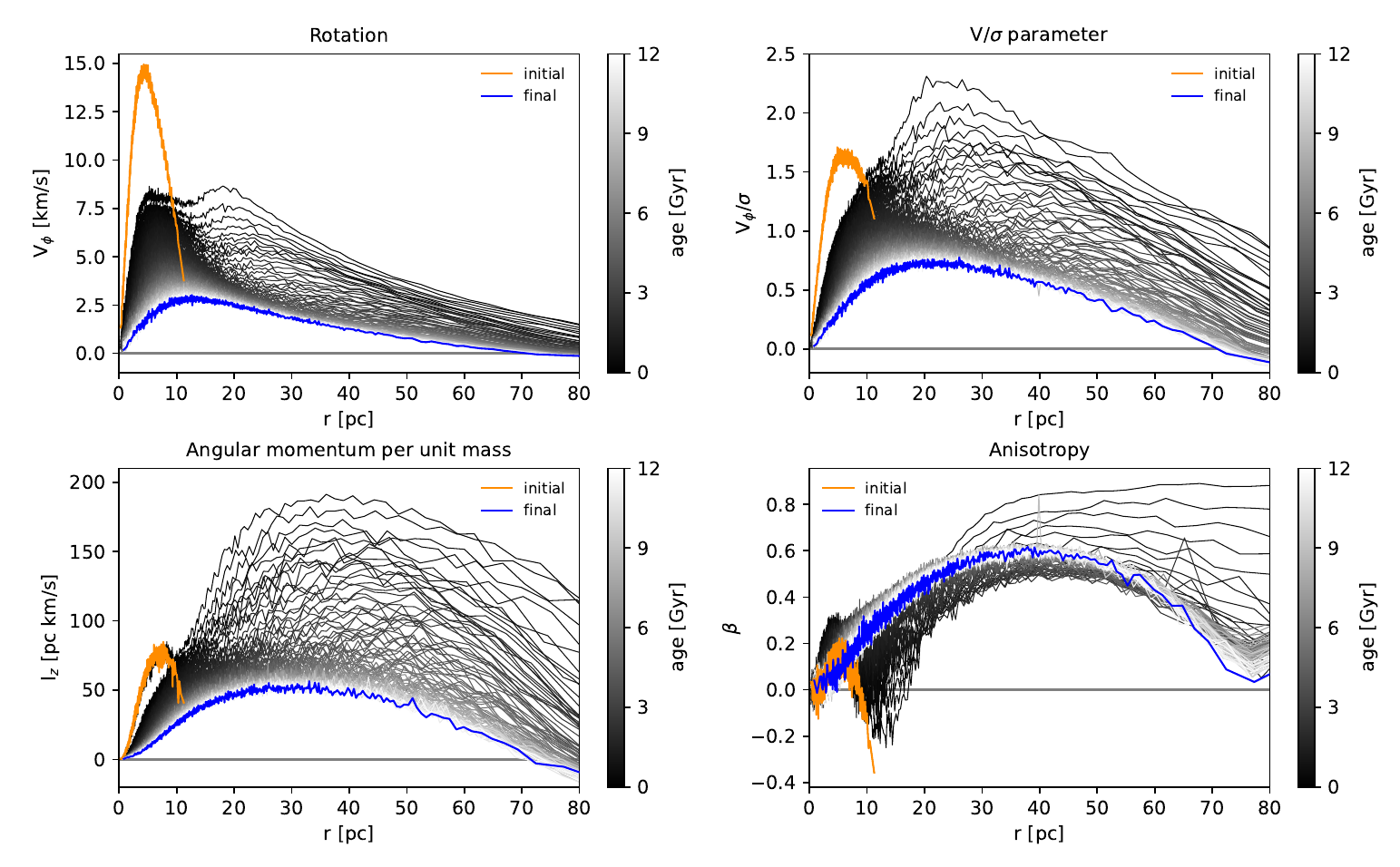}
   \caption{Evolution of the kinematic profiles of the \texttt{1.5M-A-R4-10} simulation from time t=0 Gyr (orange lines) to t=12 Gyr (blue lines). The intermediate time snapshots are color coded in gray scale. The panels report the rotation profile $V_\phi$(r), the $V_\phi/\sigma$(r) profile indicating the ratio between ordered and random motion, the z-component of the angular momentum per unit mass $l_z(r)$, and the anisotropy profile $\beta(r)$. The time resolution between the plotted lines is $\approx64$ Myr.}
              \label{1.5Mprofiles}%
    \end{figure*}
%--------------------------------------------%

\section{Evolution of GC properties}
%--------------------------------------------%
\label{sec:3}
In this section, we present the evolution of our simulated GCs. We first focus on their early evolution connected to the gravogyro-gravothermal collapse and then on the long-term evolution of their kinematic radial profiles. Finally, we discuss the evolution of global quantities as a function of relevant structural properties. 

%-----------------------------------------------
\subsection{Early evolution}
%-------------------------------------- 

The evolution of GCs is strongly shaped by collisional dynamics causing energy exchanges between stars which, due to the negative heat capacity of self-gravitating stellar systems, lead to core collapse, a rapid increase of the central density of a cluster, which can be halted by the injection of energy due to  binary formation (e.g., \citealp{Henon1961,Lynden-BellWood1968,MeylanHeggie1997}), in particular black hole binaries \citep{BreenHeggie2013}. Core-collapsed clusters are usually identified observationally via the presence of a central density cusp in their surface brightness profiles (e.g., \citealp{DjorgovskiKing1986}); but, more recently, identification methods based on kinematic data have also been put forward (e.g., \citealp{Bianchini2018,Libralato2018}). However, in the early stages of GC evolution an increase of the central density can also take place due to the quick segregation of massive objects, including BHs (see, e.g., \citealp{Kamlah2022b}); this process is referred to as the initial collapse of the core. From a general point of view, the collapse of the core in a GC can be thought of as the outcome of the gravothermal instability that collisional stellar systems experience. It has been extensively studied in GCs under various conditions (e.g., \citealp{Breen2017,Pavlik2024}) and it has been noted that the presence of internal rotation adds further features to the collapse of the core. This behavior has often been interpreted as due to an additional catastrophic process, known as gravogyro instability, which is caused by the transport and redistribution of angular momentum within a cluster (see \citealp{AkiyamaSugimoto1989,Ernst2007,Kamlah2022b} and references therein).

In order to study the onset of core collapse in our simulations, we analyze the evolution of Lagrangian radii, in particular the radii containing 50\%, 10\%, and 1\%  of the total mass, $r_{50\%}$, $r_{10\%}$ and $r_{1\%}$, respectively. The analysis carried out here assumes spherical symmetry, an assumption that is not strictly valid in these early phases of evolution where deviations from sphericity can be severe due to the onset of dynamical instabilities. This point will be discussed in the next paper of the series.

In the first row of Fig. \ref{core_collapse}, we show the evolution of the Lagrangian radii for three selected simulations with 500k stars, representative of different initial densities and rotation strengths. The first 2 models, \texttt{500k-A-R2-10} and \texttt{500k-A-R4-10}, are characterized by the same rotational support (V$_{peak}/\sigma_0=1.20$) and different densities (11.5 and 2.9 M$_\odot$\,pc$^{-2}$). The third model, \texttt{500k-C-R4-10}, is characterized by weaker rotation (V$_{peak}/\sigma_0=0.71$) and low density (2.8 M$_\odot$\,pc$^{-2}$, comparable to \texttt{500k-A-R4-10} model). The figures for all simulations are reported in Appendix \ref{appA}. 

Figure \ref{core_collapse} shows the expansion of the intermediate and outer layers (10\% and 50\% Lagrangian radii) and the contraction of the central region within $r_{1\%}$. The minimum value of $r_{1\%}$, corresponding to the maximum value of the central density, is reached by all simulations within the first 500 Myr of evolution. Higher density clusters show an earlier and deeper collapse, whereas low rotating clusters show a weaker signature of collapse. These results are in agreement with what was shown by \cite{Kamlah2022b} for rotating star clusters with $10^5$ stars. After the time of core collapse, the $r_{1\%}$ expands, as a result of heating mechanisms, such as binary formation or stellar evolution mass loss.

The second row of Fig. \ref{core_collapse} shows the evolution of the mean stellar mass contained in each Lagrangian radii. At time zero, the mean mass is the same throughout the radial extent of the clusters, indicating homogeneous non mass-segregated initial conditions. While time evolves, massive stars segregate in the center, and the mean stellar mass contained within $r_{1\%}$ reaches values of 4-6 $M_\odot$. These high values indicate the presence of massive objects, mostly BHs (and their progenitors) formed in the first few million years through SNe. Interestingly, the simulations with stronger rotations reach the maximum of mean mass around core collapse, while the simulations with lower rotation strengths reach a peak later on, beyond 1 Gyr of evolution. This trend indicates that rotation has a key role in establishing the level and the timescale of mass segregation that a cluster can reach dynamically, which can then influence the timescale in which BHs segregate in a GC center. Finally, we note that after peaking, the mean-mass declines, indicating that massive objects which sank in the center are dynamically heating and ejecting each other out of the cluster through repeated 2- and 3-body interactions (see also \citealp{BreenHeggie2013}). A much milder decline is instead observable for the mean mass within $r_{50\%}$, mostly reflecting stellar mass loss due to stellar evolution on a larger time-scale.

%-------------------------------------- Two column figure (place early!)
   \begin{figure*}
   \centering
   \includegraphics[width=1\textwidth]{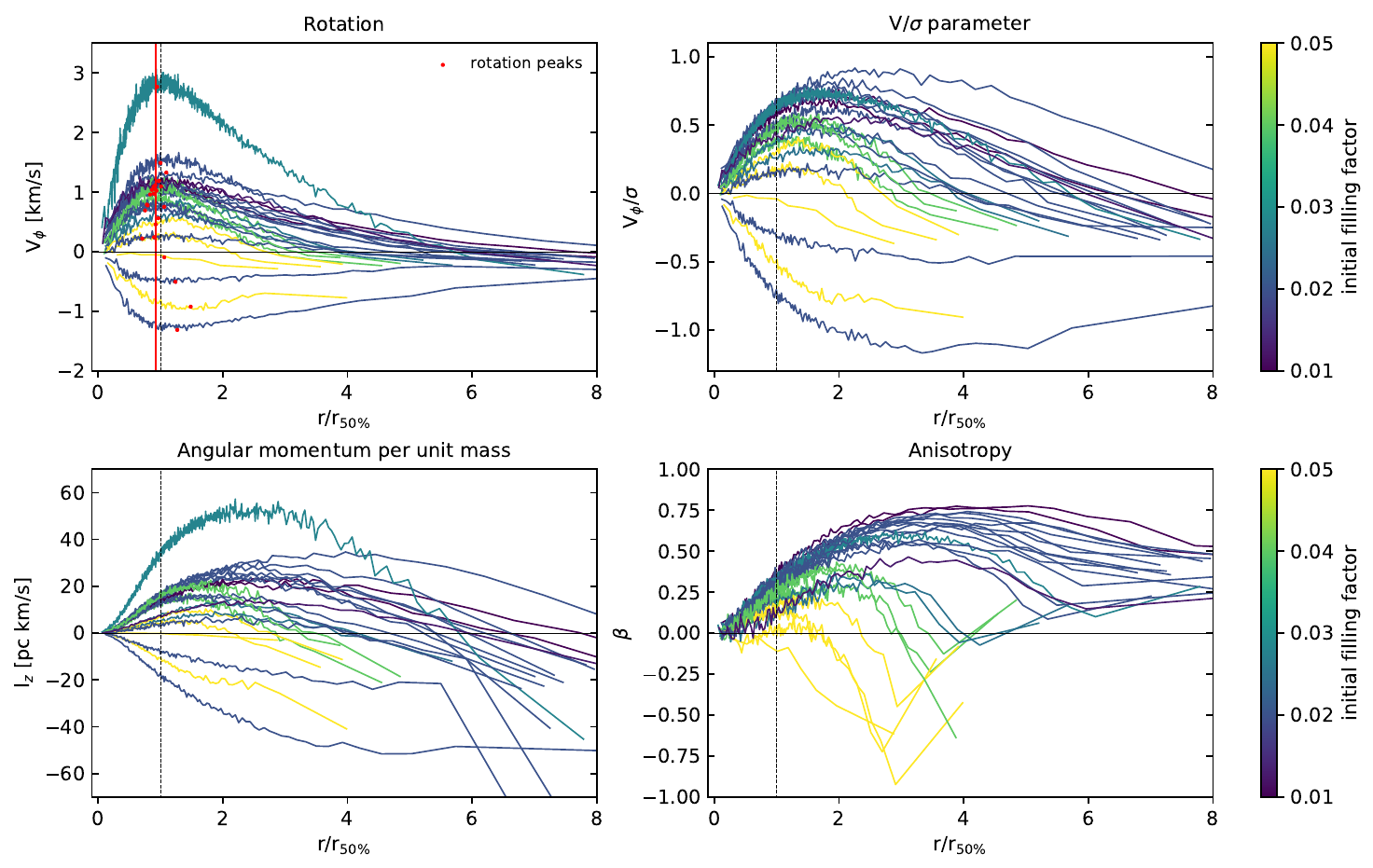}
   \caption{Kinematic profiles, as in Fig. \ref{1.5Mprofiles}, for the entire set of simulations at 12 Gyr. The profiles have been normalized by the respective half-mass radii $r_{50\%}$ (vertical dashed lines) and are color coded by the value of their initial tidal filling factor ($r_{50\%}/r_j$) tracing the tidal influence on a GC. In the rotation profile, $V_\phi$(r), we indicate the rotation peaks with red dots and their average radial position with a vertical red line.}
              \label{profs12Gyr}%
    \end{figure*}
%--------------------------------------------%
%-------------------------------------- Two column figure (place early!)
   \begin{figure*}
   \centering
   \includegraphics[width=1\textwidth]{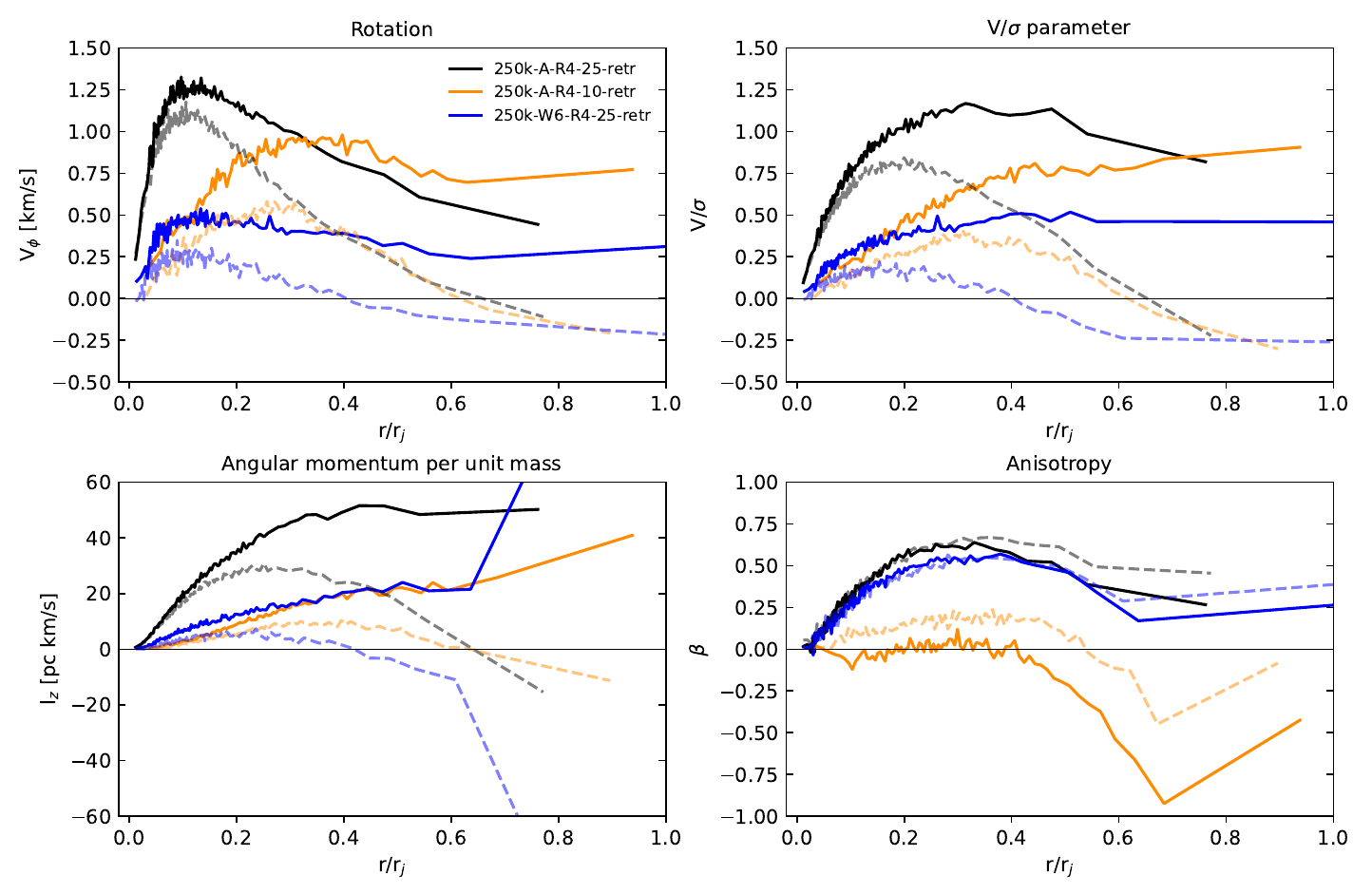}
   \caption{Kinematic profiles at 12 Gyr of three simulations with prograde rotation (with respect to the orbital angular momentum) compared to their retrograde counterparts. For the retrograde profiles, we plot the absolute values. Each profile has been normalized by its corresponding Jacobi radius, $r_j$. Retrograde GCs are indicated as solid lines, while prograde ones are shown as dotted lines. Although prograde and retrograde simulations pairs have identical initial structures and rotation strengths, they significantly differ after 12 Gyr of evolution due to the complex coupling of internal rotation and the tidal field. Retrograde GCs retain a higher amount of angular momentum and show significant differences in rotation both in their outskirts and in the intermediate regions ($\sim0.5$ $r_j$).
   }
              \label{fig:retr}%
    \end{figure*}
%--------------------------------------------%

%-------------------------------------- Two column figure (place early!)
   \begin{figure*}
   \centering
   \includegraphics[width=1\textwidth]{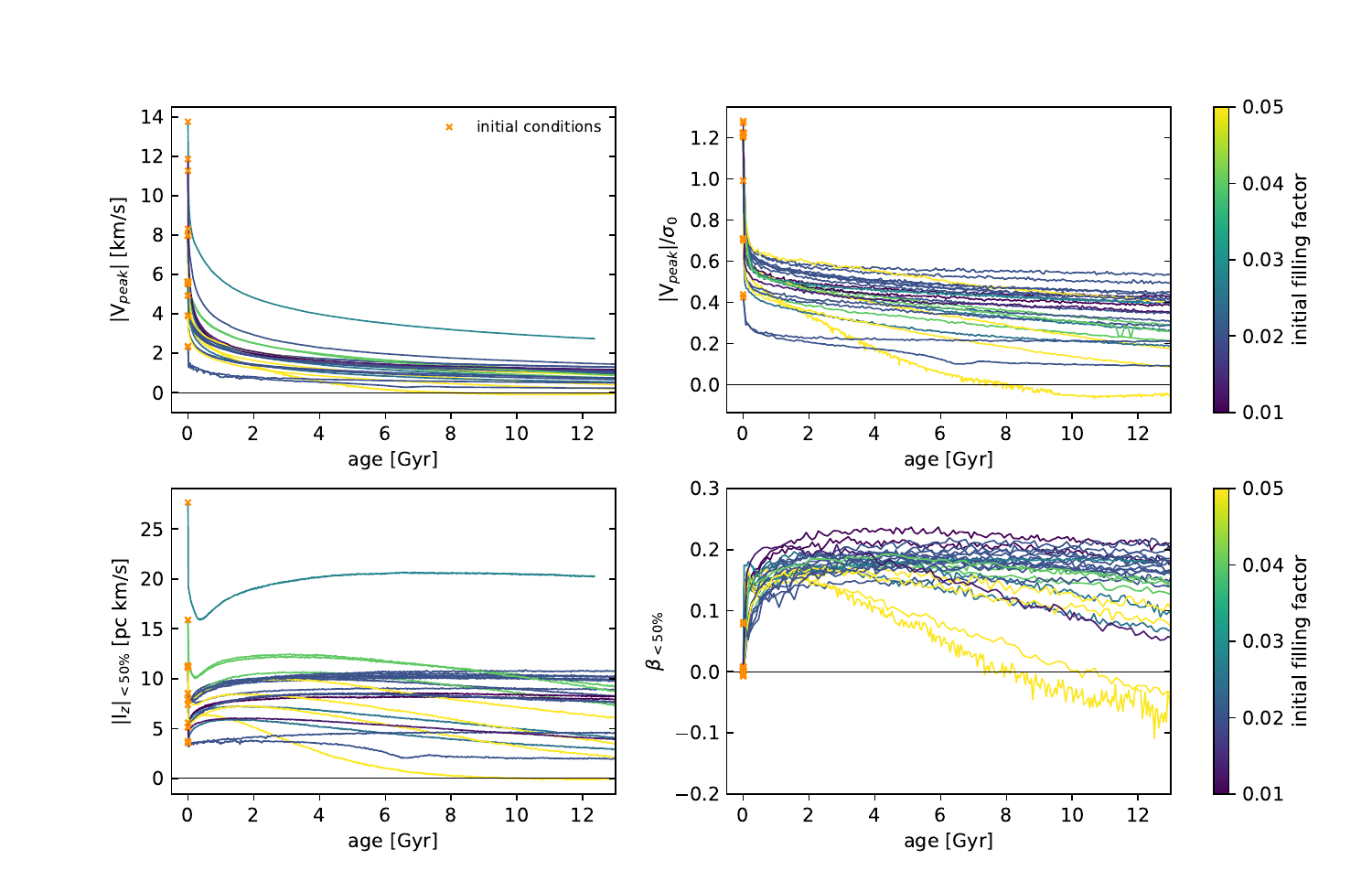}
   \caption{Time evolution of the global properties of all our simulations: absolute value of the peak of the rotation profile $V_{peak}$, $|V_{peak}|/\sigma_0$ parameter with $\sigma_0$ (the central velocity dispersion), absolute value of the z-component of the angular momentum per unit mass calculated within the half-mass radius, and anisotropy parameter $\beta$ within the half-mass radius. The initial values are indicated with orange crosses, and the curves are color coded according to their initial tidal filling factor, as in Fig. \ref{profs12Gyr}.
   }
              \label{fig:glob_t_ev}%
    \end{figure*}
%--------------------------------------------% 

\subsection{Long-term evolution of kinematic profiles}
\label{sec:long_term}
Using a spherical coordinate system $(r,\theta\,\phi)$, we computed the radial profiles of four kinematic quantities:
\begin{enumerate}
    \item the rotation profile $V_\phi(r)$, calculated as the mean of the $\phi$-component of the velocity vector;
    \item the $V_\phi/\sigma(r)$ profile, with $\sigma$ as the 1D velocity dispersion profile tracing the ratio between ordered and random motions;
    \item the z-component of the angular momentum per unit mass $l_z(r)$;
    \item the anisotropy profile $\beta(r)=1-(\sigma^2_\theta+\sigma^2_\phi)/2\sigma^2_r$, with $\beta>0$ indicating radial anisotropy.
\end{enumerate}
All profiles were calculated by dividing a GC in spherical shells such that each bin contains an equal number of stars. Figure \ref{1.5Mprofiles} reports the time evolution of such profiles for the \texttt{1.5M-A-R4-10} simulation. Each curve is color-coded with a gray scale from 0 to 12 Gyr (from back to white), and the initial profiles are indicated with orange lines and the final with blue lines.

\begin{figure*}
   \centering
   \includegraphics[width=1\textwidth]{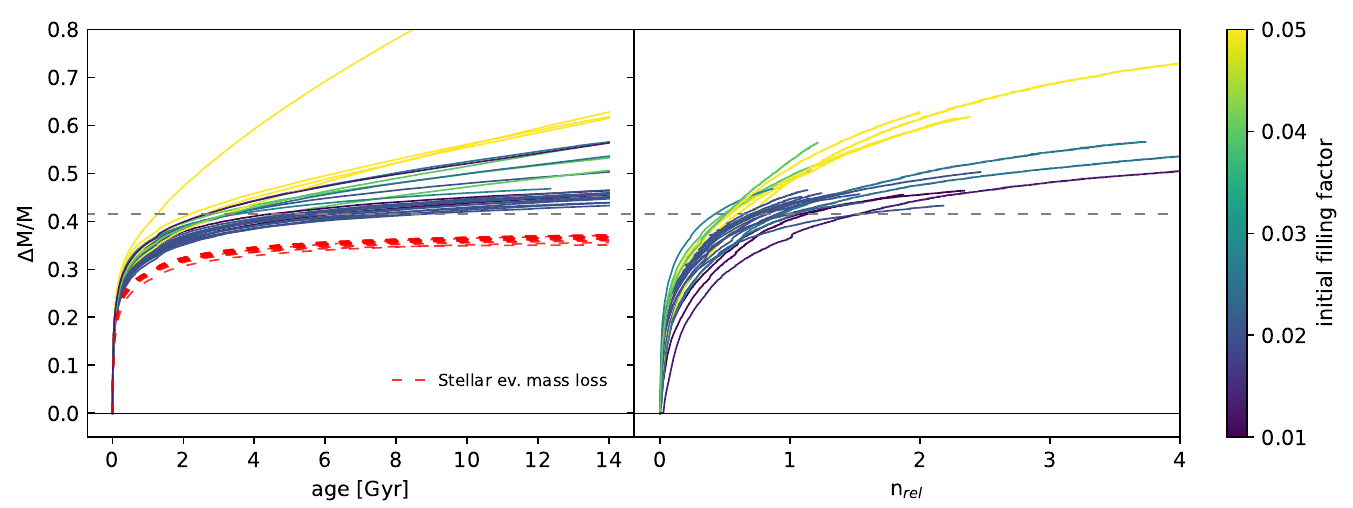}
   \caption{\textit{Left panel:} Globular cluster mass loss $\Delta M/M$ vs. age. \textit{Right panel:} Globular cluster mass loss $\Delta M/M$ vs. number of relaxation times experienced by the GC, $n_{rel}=\rm{age}/t_{rh}$. Each GC curve is color coded according to the initial tidal filling factor. The red lines indicate the mass loss due to stellar evolution alone, common for all simulations and dominating in the first gigayear of evolution. The horizontal dashed line indicates the point where the mass loss starts to be dominated by tidal effects (stellar escapers) rather than stellar evolution processes. The plot shows the importance of the initial filling condition in defining the amount of mass loss cumulatively experienced by a GC after 12 Gyr of evolution. The $n_{rel}$ plot clearly shows that GCs formed with initial filling factors higher than 0.035 lose significantly more mass given the same relaxation condition.
   }
              \label{fig:massloss_colorcoded}%
    \end{figure*}

By analyzing the differences between the initial and final profiles of Fig. \ref{1.5Mprofiles}, we can appreciate the effects of the long-term evolution of GCs. In particular, we note an expansion of the cluster and a loss of angular momentum, clearly visible in the $V_\phi(r)$, $V_\phi/\sigma(r)$, and $l_z(r)$ profiles as a reduction of the respective peaks and a shift toward larger radii. The anisotropy profile $\beta(r)$ shows that the cluster starts with isotropy in the center, mild radial anisotropy in the intermediate regions, and mild tangential anisotropy in the outskirts. The long-term evolution imprints a similar trend to the $\beta(r)$ profile at 12 Gyr, but with an enhanced radial anisotropy that builds up with time (consequence of the radial expansion of the cluster). Moreover, we note that the evolution of the kinematic profiles is more prominent in the early times compared to the much smoother evolution occurring on a longer timescale. This is due to the changes induced by early mass loss driven by stellar evolution, most effective in the early epochs of evolution (see also Sect. \ref{sec:global_trends} and Fig. \ref{fig:massloss_colorcoded}).

These general features described for the \texttt{1.5M-A-R4-10} simulation also apply to the entire set of simulations. However, the large parameter space covered by the initial conditions does imprint a remarkable variety of properties. In Fig. \ref{profs12Gyr} we report the kinematic profiles of all simulations at 12 Gyr, normalized by their respective half-mass radius. Since we are interested in unveiling the imprint of the initial conditions on the final state of a GC, we color code each profile according to their initial tidal filling factor, defined as the ratio between the half-mass radius and the Jacobi radius, $r_{50\%}/r_j$. This quantity provides both an estimate of the initial density of a cluster and a measure of the intensity of the tidal effects that a cluster experiences due to the presence of the external environment (a higher filling factor means that a GC is more vulnerable to tidal effects).

The first panel of Fig. \ref{profs12Gyr} shows the variety of rotation profiles $V_\phi(r)$ produced by the set of simulations. The three negative rotation profiles correspond to clusters with retrograde internal rotation with respect to their orbital motion around the galaxy (we refer the reader to Sect. \ref{sec:retr} for a discussion of the differences between prograde and retrograde clusters). For each profile, we identified the rotation peak by performing a fit of the profiles with the following relation routinely used to model rotation profiles in GCs (see, e.g., \citealp{Mackey2013,Kacharov2014,Bianchini2018b}):
\begin{equation}
V(r)=\frac{2V_{peak}}{R_{peak}}\frac{r}{1+(r/R_{peak})^2},
\end{equation}
where $V_{peak}$ and $R_{peak}$ represent the peak of the rotation curve and its radial position.

In Fig. \ref{profs12Gyr} we indicate the rotation peaks with red points and the average of their radial position with a red vertical line. The profiles are characterized by a wide range of rotation amplitudes (from 0.2 to 3 km/s) and, noticeably, all rotation peaks are located around the half-mass radius $r_{50\%}$ (as demonstrated by the red vertical line almost coinciding with $r_{50\%}$). Interestingly, the clusters with retrograde rotation have flatter rotation profiles, with peaks beyond the half-mass radius (see also Sect. \ref{sec:retr}). The shapes and amplitudes resulting from the long-term evolution of our simulations are fully consistent with observations of today's MW GCs (e.g., \citealp{Bianchini2018b}).  

As in the case of the rotation profiles, the $V_\phi/\sigma(r)$ and $l_z(r)$ profiles also show an analogous variety of shapes. We note here that the three types of kinematic profiles that trace rotation in GCs do not show a sharp dependence on the initial filling factor. However, GCs with higher initial filling factors tend to show lower rotation signals, consistent with a stronger loss of angular momentum driven by the tidal field. We discuss evidence of this trend further in Sect. \ref{sec:global_trends}. 

Of all kinematic profiles in Fig. \ref{profs12Gyr}, the anisotropy profiles $\beta(r)$ show the widest diversity of shapes. In particular, despite the fact that all simulations start with a similar and globally weak radial anisotropy, different anisotropy flavors build up over time. All profiles are characterized by isotropy in the central region, but in the intermediate and outer parts they can exhibit: (i) strong radial anisotropy, (ii) mild radial anisotropy with quasi-isotropy in the outskirts, (iii) mild radial anisotropy with tangentiality in the outskirts, or (iv) fully tangential anisotropy. All these different flavors of anisotropy show a clear correlation with the initial filling factor, as is evident from the color coding: GCs with initially higher filling factors develop more tangential anisotropic profiles, while GCs with more underfilling initial conditions develop stronger radial anisotropy.

From a theoretical perspective, dynamical evolution, combined with mass-loss driven by stellar evolution, can naturally imprint radial anisotropy, due to the ejection of stars from the core to the halo of a cluster, a phenomenon that is particularly efficient during the post-core collapse phase \citep{SpitzerShapiro1972,GierszHeggie1996}. In contrast, the presence of a tidal field can quench radial anisotropy, bringing the outer part of the system toward isotropy and tangential anisotropy \citep{GierszHeggie1997}, due to the preferential stripping of stars in radial orbits (e.g., \citealp{TakahashiLee2000,BaumgardtMakino2003}).
These effects, due to the coupling of the tidal field, internal dynamics, and stellar evolution, have been explored in previous simulations clearly showing that GCs can develop tangential anisotropy if they experience strong tidal fields (\citealp{Tiongco2016b,Zocchi2016,Bianchini2017}). Our simulations confirm these results and extend their validity to the mass regime of real GCs with $>10^5$ stars, while also including realistic stellar evolution prescriptions.

Before concluding this section, a particular case worth mentioning is the simulation \texttt{250k-A-R2-5} characterized at 12 Gyr by an extremely low rotation signal. This model starts as a compact and fast rotating GC ($r_{50\%}=2$ pc and $V_{peak}=8.31$ km/s) and experiences a particularly strong mass loss of $\Delta M /M=0.92$ (see properties in Tables \ref{table:1} and \ref{table:2}). This is the result of the interplay between the initial high concentration of the cluster, the strong tidal field that the model experiences (due to the proximity of its orbit to the center of the galaxy, d=5 kpc), and the escape of stellar remnants due to their natal kick. These three ingredients trigger a strong mass loss that can lead to an abrupt cluster dissolution (a behavior often referred as "jumping"; see, e.g., \citealp{Contenta2015,Giersz2019}). The cluster quickly expands, becomes strongly tidally filling and it loses substantial mass, such that its $r_{50\%}$ starts to decrease.
This strong mass loss leads to a significant loss of angular momentum, erasing any original rotation features, which explains the almost constant rotation profile with $V_{\phi}\approx0$ km/s within the half-mass radius and the slightly negative rotation in the outer parts. The rotation profile outside the half-mass radius has a solid-body pattern with angular velocity ($\omega=28.03$ km/s/kpc) approximately equal to half of the angular velocity of the cluster orbital motion around the host galaxy ($\Omega=58.66$ km/s/kpc). This is fully consistent with the results of \cite{Tiongco2016a}, predicting that a tidal field can imprint an internal rotation signature that is partially synchronized with the orbital angular momentum ($\Omega\approx-\frac{1}{2}\omega$). Finally, the same model shows a strong signature of tangential anisotropy already within the half-mass radius. This specific case is rather instructive, as it shows how the observation of a lack of rotation in a cluster (or of a mild solid body rotation), combined with a tangential anisotropy signature, can indicate that the stellar system has undergone significant mass loss and is close to disruption.

%-------------------------------------- Two column figure (place early!)
%--------------------------------------------%

\subsection{The case of retrograde rotation}
\label{sec:retr}
Our suite of models contains three pairs of GCs with identical structural initial conditions but different orientations of the rotation axis. Specifically, three simulations have prograde internal rotation with respect to the orbital angular momentum (\texttt{250k-A-R4-25}, \texttt{250k-A-R4-10}, and \texttt{250k-W6-R4-25}) and three simulations have retrograde rotation (\texttt{250k-A-R4-25-retr}, \texttt{250k-A-R4-10-retr}, and \texttt{250k-W6-R4-25-retr}). As highlighted in previous works, the tidal field will act differently in the prograde and retrograde cases: stars in prograde orbits are energetically less stable and more prone to be stripped by the tidal field; this leaves an excess of stars in retrograde orbits in a given cluster (see, e.g., \citealp{Keenan1975,FukushigeHeggie2000,Ernst2008,Tiongco2016a,Tiongco2017,Claydon2019}). As a consequence, the amount of angular momentum lost by a cluster will depend on the interplay between internal rotation and tidal field, and it will ultimately shape the rotation curves of present-day GCs. 

Figure \ref{fig:retr} shows the kinematic profiles of these simulations, with dashed lines indicating prograde GCs and solid lines retrograde ones. To facilitate the comparison, we plot the absolute values for the retrograde profiles. We normalized the profiles by the Jacobi radius $r_j$ (the limiting radius of the GCs) to highlight whether the differences affect only the outer regions, where the tidal field is most effective, or also the inner and intermediate parts. We stress that for each pair all the kinematic profiles at t=0 Gyr are identical, so the differences in the plots can only be attributed to the prograde/retrograde configurations. Four main significant differences are noticeable:
\begin{enumerate}
    \item Retrograde GCs are able to retain a higher amount of rotation. This is clearly visible by the overall higher values of rotation and the corresponding higher peaks in the profiles. 
    \item The $V/\sigma$ and $l_z$ profiles of retrograde GCs are characterized by a flat (or slightly increasing) shape in the intermediate/outer regions, whereas prograde GCs show a peak (around 0.2-0.4 $r_j$) followed by a decrease. 
    \item Prograde GCs show counter rotation signatures (with respect to the inner and intermediate regions) for radii >0.5-0.7 $r_j$, contrary to retrograde GCs where the rotation pattern keeps the same direction along the entire radial extent. These counter-rotating signatures appear in all prograde models, as clearly visible in Fig. \ref{profs12Gyr} and are consistent with tidal field induced internal rotation, as described in Sect. \ref{sec:long_term} for the \texttt{250k-A-R2-5} model.
    \item Differences in anisotropy between retrograde and prograde GCs are small, except in the outer regions, where retrograde GCs are more isotropic and can display tangentiality, specifically in the case of stronger tidal fields.
\end{enumerate}

These results highlight the importance of the coupling between tidal field and internal rotation in shaping rotation signatures and indicate that the properties of kinematic profiles provide a powerful diagnostic to distinguish prograde from retrograde GCs. In particular, our work shows that the differences between prograde and retrograde GCs are already detectable in the intermediate regions of the cluster, at distances $\sim0.5\,r_j$, rather than exclusively in the outermost parts, where observational analyses are complicated by low number statistics and high contamination from background and foreground stars. We note, however, that real clusters may be subject to
more complex configurations between their orbital and internal rotation inclinations.

%-------------------------------------- Two column figure (place early!)
   \begin{figure*}
   \centering

   \includegraphics[width=1\textwidth]{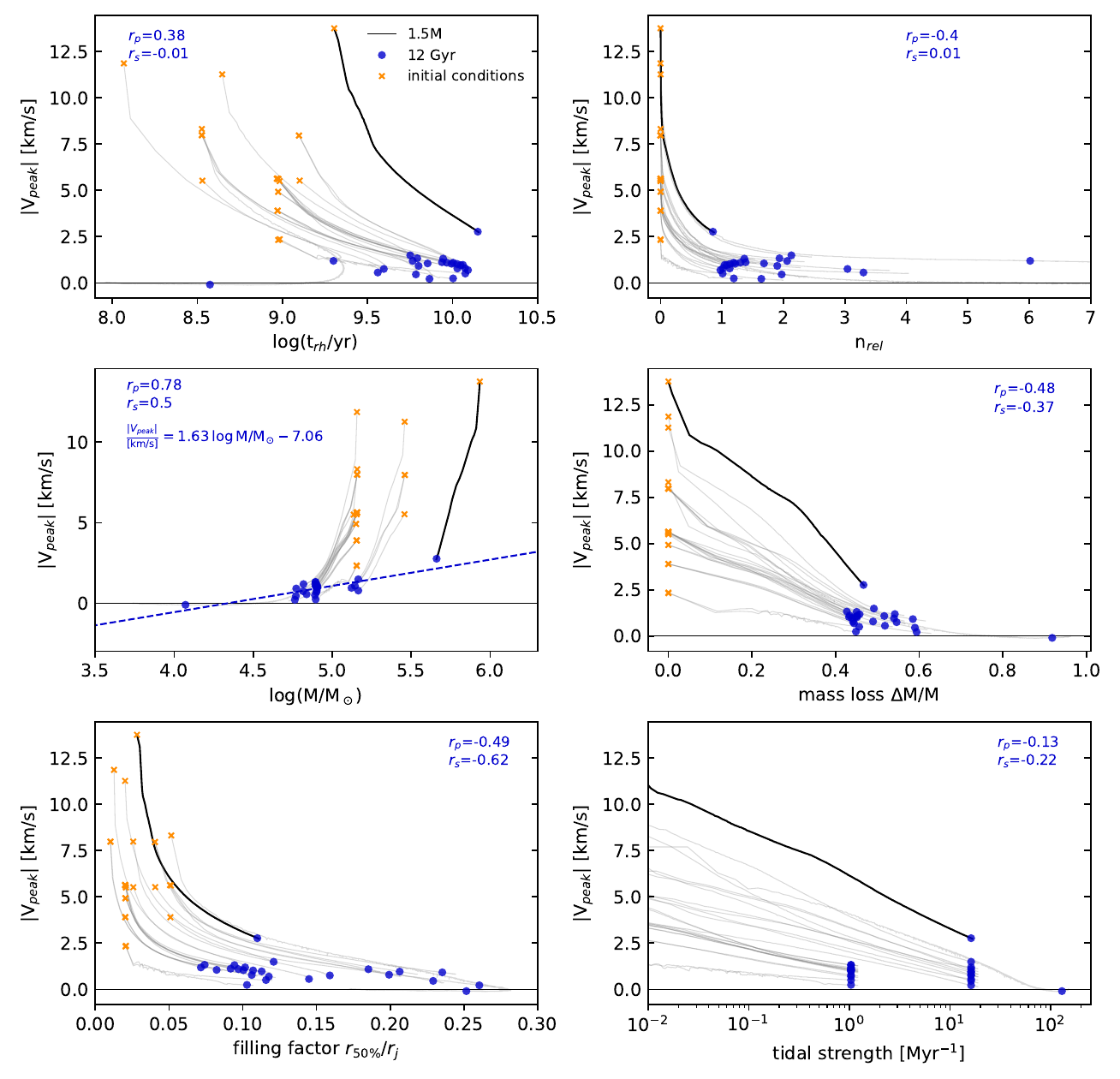}
   \caption{Correlation between the peak of the velocity profiles $|V_{peak}|$ of all simulations versus various quantities: the relaxation time $t_{rh}$ at a given snapshot; the number of relaxation times $n_{rel}$; the total mass $M$; the mass loss $\Delta M/M$; the tidal filling factor $r_{50\%}/r_t$; and the cumulative tidal strength experienced by a cluster (defined in Sect. \ref{sec:global_trends}). Orange crosses indicate the initial conditions, blue circles the snapshots at 12 Gyr, and the gray lines the evolutionary tracks. We highlight in black the track corresponding to the simulation \texttt{1.5M-A-R4-10}. For each panel we performed a Pearson and a Spearman correlation test for the data points at 12 Gyr, and we indicate the corresponding coefficients ($r_p$ and $r_s$), and plot the corresponding linear fit (dotted blue line) for the panel showing the strongest correlation, $|V_{peak}|$ vs. M.}
              \label{fig:vpeak_corr}%
    \end{figure*}
%--------------------------------------------%
%-------------------------------------- Two column figure (place early!)
   \begin{figure*}
   \centering
   \includegraphics[width=1\textwidth]{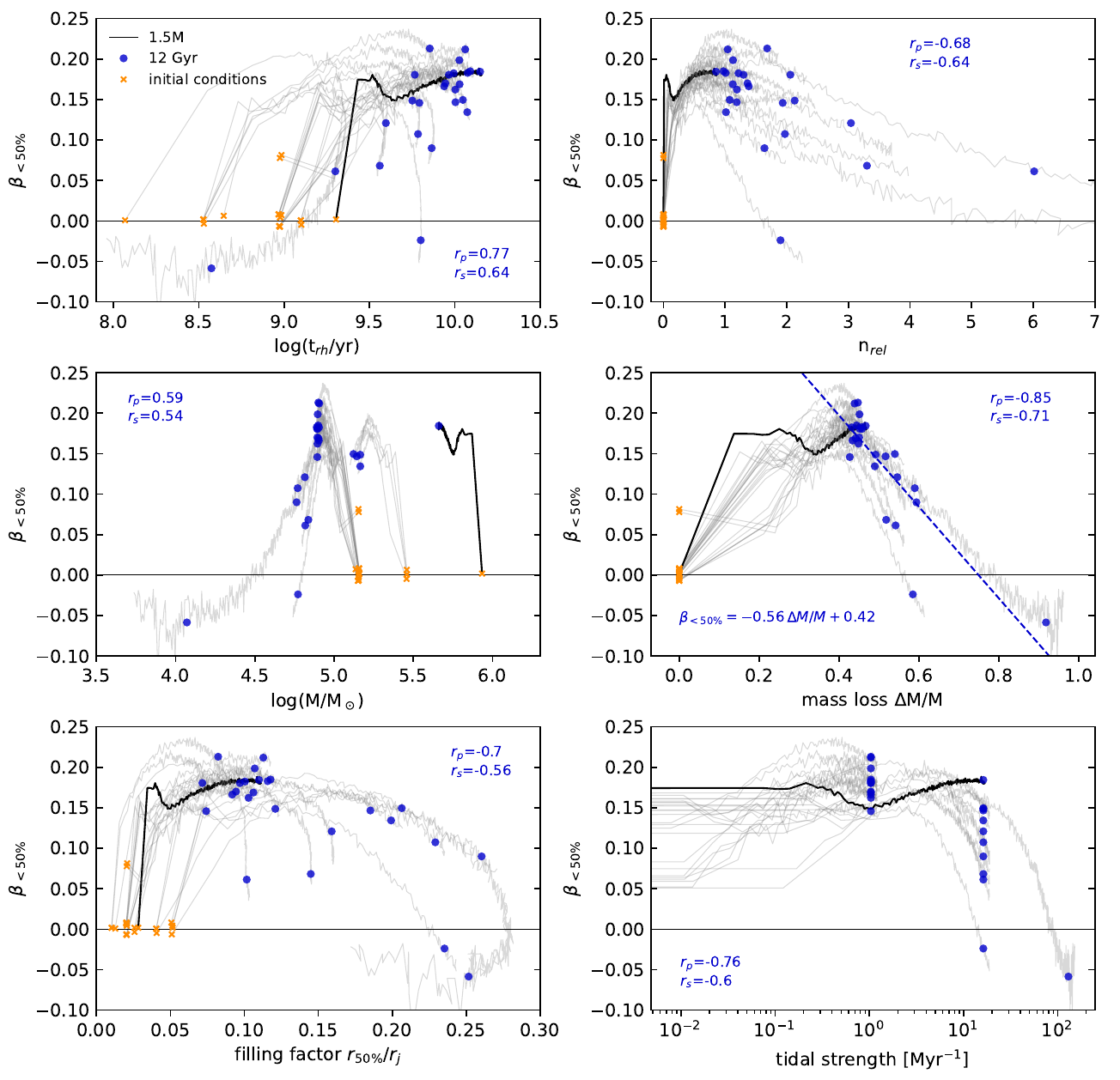}
   \caption{Correlation between the anisotropy $\beta_{<50\%}$ calculated within the half-mass radius and the same quantities as in Fig. \ref{fig:vpeak_corr}. The strongest correlation at 12 Gyr was found for $\beta_{<50\%}$ vs. $\Delta M/M$.}
              \label{fig:beta_corr}%
    \end{figure*}
%--------------------------------------------%

\subsection{Long-term evolution of global kinematic quantities}
\label{sec:global_trends}
Next, we focus on characterizing the evolution of global kinematic properties, namely:
\begin{enumerate}
    \item the absolute value of the peak of the rotation profile $|V_{peak}|$;
    \item the $|V_{peak}|/\sigma_0$ parameter, with $\sigma_0$ as the central value of the 1D velocity dispersion profile;
    \item the z-component of the specific angular momentum within the half-mass radius $r_{50\%}$;
    \item the velocity anisotropy within the half-mass radius $r_{50\%}$, $\beta_{<50\%}$.
\end{enumerate}

Figure \ref{fig:glob_t_ev} presents the evolution of these four global properties as a function of age. Each simulation is color-coded according to its initial filling factor, as in Fig. \ref{profs12Gyr}. As already noticed for the kinematic profiles, models with a higher initial filling factors show a more rapid change of their kinematic properties in the long term, highlighting again the importance of the tidal field in shaping the evolution of these quantities. This is particularly evident for the velocity anisotropy evolution, as it was also the case for Fig. \ref{profs12Gyr}. However, the most rapid and significant changes occur at earlier times ($\lesssim1$ Gyr), when stellar evolution processes are the most efficient and induce a mass loss of up to 30\% to all GCs, as shown in Fig. \ref{fig:massloss_colorcoded}. The mass loss rate due to stellar evolution alone is dominant up to an average mass loss of 0.42 (see horizontal line of Fig. \ref{fig:massloss_colorcoded}): when the clusters reach this total mass loss value, the main mechanism driving mass loss is the tidal escape of stars. \footnote{This is calculated by measuring the mass loss rate for each model, separately for stellar evolution and for stellar escapers. The mass loss rate due to stellar escapers becomes dominant when the clusters have experienced a total mass loss of $\Delta M/M \sim0.37-0.43$.} The time at which tidal effects become dominant is different for each model, depending on the tidal field strength they experience, and it ranges from 800 Myr to $\sim9.5$ Gyr.

The difference in mass loss due to the initial tidal conditions of a cluster is noticeable in the right panel of Fig. \ref{fig:massloss_colorcoded}, where mass loss is plotted against the relaxation state of a cluster $n_{rel}$, defined as the number of relaxation times a cluster has experienced, $n_{rel}$=t$_{age}$/t$_{rh}$ (with t$_{rh}$ representing the relaxation time of a given time snapshot). After a common initial evolution, GCs with similar $n_{rel}$ (i.e., similar internal dynamical conditions) lose mass (and stars) more efficiently if they were born with a higher filling factor, specifically $r_{50\%}/r_t\gtrsim0.035$.

In the case of the evolution of the specific angular momentum (Fig. \ref{fig:glob_t_ev}), we note that most of the simulations reach a minimum around the time of core collapse (see Figs. \ref{fig:cc_rad} and \ref{fig:cc_mass}). This reflects the behavior of the spatial distribution of a cluster that reaches a maximum concentration (i.e., minimum radial extent of the core region) at core collapse.   
Finally, we note that the different natal kicks prescriptions used in our suite of models (see Sect. \ref{sec:initial_conditions}) have only a minor effect on the kinematic quantities analyzed here. Table \ref{table:2} shows that simulations with different natal kicks experience similar mass loss at 12 Gyr, and the models characterized by lower kicks display marginally lower rotational support, marginally higher radial anisotropy and are more spatially extended. These differences can be attributed to the different number of stellar remnants retained by a cluster depending on the natal kick prescription. We will analyze this topic in more depth in a follow up paper.

In the last part of this section, we focus on two main kinematic properties, |V$_{peak}$| and $\beta_{<50\%}$, to identify the physical ingredients that primarily shape their evolution. In Fig. \ref{fig:vpeak_corr} and \ref{fig:beta_corr}, we plot the rotation and anisotropy parameters against the half-mass relaxation time t$_{rh}$, $n_{rel}$, the total mass of the system, the mass loss, the filling factor, and the cumulative tidal strength experienced by a cluster.\footnote{The cumulative tidal strength is defined as in \cite{Bianchini2017}, by considering the leading eigenvalue of the diagonalized tidal tensor \citep{Renaud2011}. In case of circular orbits with radius $R_g$ around a point-mass galaxy $M_g$, this corresponds to $3\,G\,M_g/{R_g}^3\,t_{age}$. The higher this parameter, the stronger the tidal field experienced by a cluster.} All of these quantities are measured for every snapshot. In each plot, the evolutionary tracks are shown as gray lines and the initial and 12 Gyr snapshots with orange and blue symbols, respectively. For the data points at 12 Gyr we perform a Pearson and Spearman correlation test, and indicate the corresponding coefficients ($r_p$ and $r_s$) in each panel.

In the case of rotation (Fig. \ref{fig:vpeak_corr}), the strongest correlation was obtained for |V$_{peak}$| versus log(M/M$_{\odot}$), with $r_p=0.78$ and $r_s=0.5$. The corresponding linear fit of the data is plotted as a dotted line and gives the following relation:
\begin{equation}
   \frac{ |V_{peak}| }{ [\rm{km\,s^{-1}}]} = 1.63\, \log M/M_{\odot} - 7.06.
\end{equation}
 The rotation-mass correlation was previously observed for MW GCs (e.g., \citealp{Bellazzini2012,Kamann2018,Bianchini2018b,Leitinger2025}), implying that more massive clusters rotate faster. This is often taken as the result of the dissipation of angular momentum (e.g., \citealp{EinselSpurzem1999,Tiongco2017}), as more massive systems would lose angular momentum less efficiently due to their longer relaxation timescales. However, we notice that our simulations do not show a strong correlation between |V$_{peak}$| and $n_{rel}$ (or t$_{rh}$), suggesting that the observed rotation-mass relation could be due to both the formation mechanism of GCs and evolutionary processes. This would imply that more massive clusters already rotate faster at formation, as is the case in our suite of simulations\footnote{Since our initial conditions are set as self-consistent models in equilibrium, a higher total mass automatically implies a higher absolute value of rotation, for a model with a given rotational support.}. Interestingly, by normalizing V$_{peak}$ by the initial peak rotation value (i.e., factoring out the impact of the initial rotation) we obtain a milder V$_{peak}$ versus M correlation ($r_p=0.49$ and $r_s=0.35$). This supports the idea that the initial conditions have a key role in setting the rotation at later ages.
 
 In Fig. \ref{fig:vpeak_corr}, other milder (anti)correlations are present, in particular for |V$_{peak}$| versus filling factor ($r_p=-0.49$ and $r_s=-0.62$) and for |V$_{peak}$| versus mass loss $\Delta$M/M ($r_p=-0.48$ and $r_s=-0.37$). They indicate that also the tidal field has an impact in shaping rotation (a closer look at this is presented in Sect. \ref{primordial_rotation}).

In the case of velocity anisotropy (Fig. \ref{fig:beta_corr}), the strongest correlation is found between anisotropy and mass loss ($r_p=-0.85$ and $r_s=-0.71$), but also other comparable correlations are found between anisotropy and relaxation conditions (t$_{rh}$ and $n_{rel}$), the tidal field (filling factor and tidal strengths), and, to a lesser extent, the total mass. This suggests that both internal processes driven by two-body relaxation, and external processes, driven by the interaction with the host galaxy, contribute to shaping the velocity anisotropy of a cluster. The strong correlation obtained for the mass loss $\Delta$M/M, indicates that this parameter efficiently captures these complex physical processes, as it traces the stellar escapes due to the interplay between two-body relaxation and tidal-field stripping.\footnote{The total mass loss also includes stellar evolution processes which are, as shown in Fig. \ref{fig:massloss_colorcoded}, roughly the same for all simulations.} A linear fit gives the following relation: 
\begin{equation}
    \beta_{<50\%} = -0.56\, \Delta M/M + 0.42.
\end{equation}
Interestingly, all simulations reach a maximum value of radial anisotropy around a mass loss of $40\%$.

The general validity of the anisotropy and mass-loss relation is also supported by previous studies. In particular, \cite{Bianchini2017} found a similar relation for a set of nonrotating simulations, with a smaller number of stars (N=50k), evolved in a large variety of time-dependent tidal fields. Their work indicates that the transition between radially and tangentially anisotropic GCs occurs at mass loss $\simeq60\%$ (see their Fig. 4). In our case, only two simulations have tangential anisotropy ($\beta_{50\%}<0$), but the linear fit suggests that this transition happens in a similar mass-loss regime, around $60-80\%$. Finally, the observational study of \cite{Watkins2015} detected a correlation between anisotropy and relaxation time, with a transition between radial and tangential anisotropic clusters at t$_{rh}=10^9$ yr. This is fully consistent with the relation found in the first panel of Fig. \ref{fig:beta_corr} and with Fig. 2 of \cite{Bianchini2017}. 

In summary, our simulations confirm that the evolution of the kinematic properties, in particular the velocity anisotropy, is driven by mass loss, which in turn is due to the combination of internal and external dynamical processes. For internal rotation, the results indicate that the values observed at present are also dependent on the amount of rotation at formation. We discuss this point further in Sect. \ref{primordial_rotation}, where we directly link the present-day rotation to the primordial rotation.

%-------------------------------------- 
\section{Connecting present-day and primordial GC properties}
%-------------------------------------- 
\label{sec:4}
In order to connect GCs present-day properties with their properties at formation, we exploit the evolutionary tracks derived from our suite of models. In particular, we focus on reconstructing the amount of primordial internal rotation of GCs and carry out a comparison with the structural parameters of MW GCs and high-z proto-GCs. 

%-------------------------------------- Two column figure (place early!)
   \begin{figure}
   \centering
   \includegraphics[width=1\columnwidth]{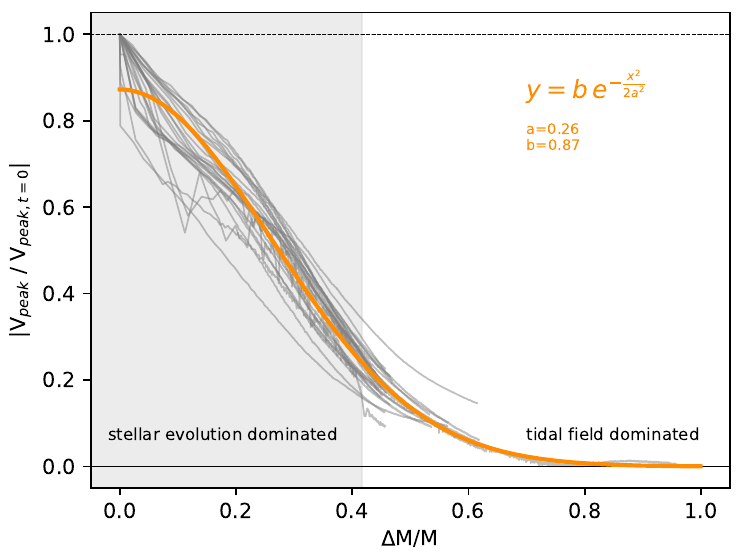}
   \caption{Evolution of the absolute value of the rotation peak $V_{peak}$ normalized by the initial value at time t=0, $V_{peak,t=0}$ as a function of mass loss $\Delta M/M$. All simulations follow a similar track with a small variation, indicating that mass loss is the key parameter driving the decrease of rotation strength. The orange line is a fit of a simple function able to reproduce the shape of this evolutionary track. By measuring $V_{peak}$ and mass loss at the present time, one can estimate the initial rotation peak (and the rotation peak at any given time). The gray shaded area indicates the phases where mass loss is dominated by stellar evolution or by tidal stripping (see also Fig. \ref{fig:massloss_colorcoded}).}
              \label{fig:reconstruct_v}%
    \end{figure}
%--------------------------------------------%
%-------------------------------------- Two column figure (place early!)
   \begin{figure*}
   \centering
   \includegraphics[width=1\textwidth]{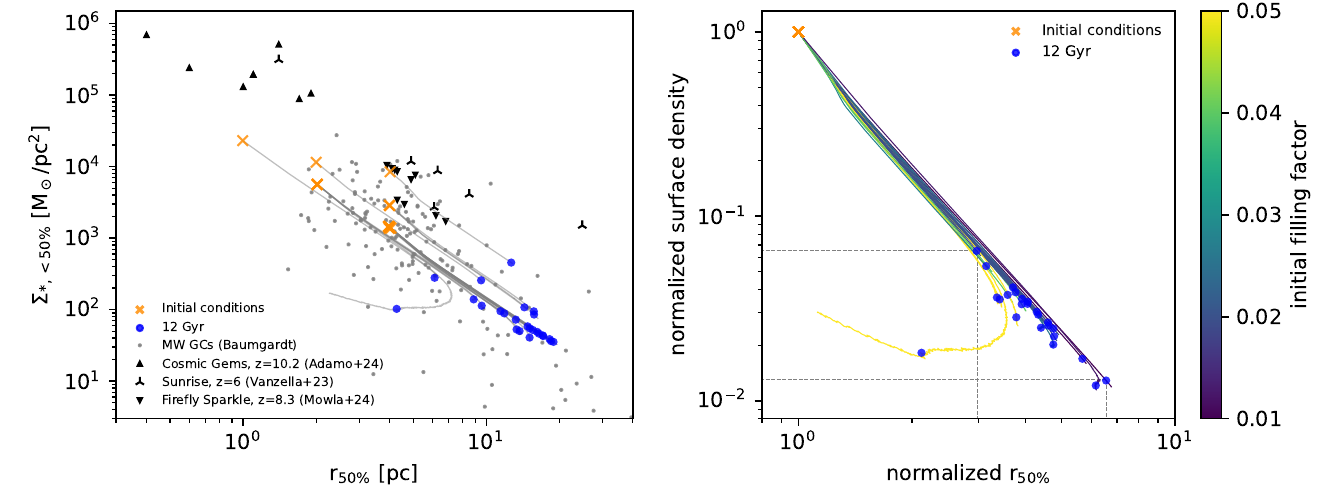}
   \caption{\textit{Left panel:} Evolution of the surface density calculated within the half-mass radius, $\Sigma_{*,<50\%}$, as a function of the half-mass radius compared to observation of the present-day MW GCs (gray dots) and high-redshift observations of proto-GCs (black symbols). Orange crosses and blue dots indicate the initial and 12 Gyr snapshots of the simulations. \textit{Right panel:} Normalized version of the left panel where each simulation is color coded according to its initial tidal filling factor. This plot shows that in this parameter space, GCs follow a similar track: After 12 Gyr of evolution, the surface density decreases by a factor of $\approx 1-6 \times 10^{-2}$, and the half-mass radius increases by a factor of $\approx 3-5$.}
              \label{fig:highz}%
    \end{figure*}
%--------------------------------------------%

\subsection{Reconstructing GC primordial rotation}
\label{primordial_rotation}
The correlation found in Sect. \ref{sec:global_trends} between rotation strength and current mass indicates that both the initial rotation and evolutionary processes cooperate in shaping this relation: GCs originally formed more massive and with stronger rotation would retain a stronger rotation after several Gyr of evolution. 

To factor out the effects of initial rotation, we plot in Fig. \ref{fig:reconstruct_v} the evolution of the normalized rotation peak $|V_{peak}/V_{peak,t=0}|$ (where $V_{peak,t=0}$ corresponds to the peak of the rotation profile at time t=0 Gyr) as a function of mass loss $\Delta M/M$. All models follow very similar evolutionary tracks, indicating that mass loss is responsible for the decrease of rotational support. A similar result was highlighted in \cite{Tiongco2017} for three smaller rotating simulations.

Figure \ref{fig:reconstruct_v} indicates that by measuring the peak of rotation at the present time and the mass loss experienced by a cluster, it is possible to reconstruct the rotation value at earlier times. For this reason, we performed a fit of these evolutionary tracks using the following equation:
\begin{equation}
    \left|\frac{V_{peak}}{V_{peak,\,t=0}}\right|= b\,\exp\left(-\frac{1}{2}\frac{(\Delta M/M)^2}{a^2}\right),
\label{eq:rot_ev}
\end{equation}
obtaining best-fit values of $a$ and $b$ of $a=0.26$ and $b=0.87$. Note that the formula is not physically motivated, but is found to provide a good description of all simulations throughout their entire long-term evolution. The plot also shows the regime where the mass loss rate of a cluster is dominated by stellar evolution processes or by stellar escapers: this transition happens when the models have lost on average a total of 42\% of their total mass (see also Sect. \ref{sec:global_trends} and Fig. \ref{fig:massloss_colorcoded}).

Figure \ref{fig:reconstruct_v} demonstrates the strong impact of the long-term evolution in shaping the rotation strength. For a GC that underwent a mass loss of 50\% after 12 Gyr of evolution (as is the typical case of our suite of simulations), the rotation amplitude has decreased by $\approx84\%$. The rotation decreases by $\approx75\%$ if the GC has experienced 40\% of mass loss. This implies that mass loss due to stellar evolution alone ($\Delta M/M\approx40\%$ throughout a cluster's life; see Sect. \ref{sec:global_trends} and Fig. \ref{fig:massloss_colorcoded}) is responsible for a reduction in rotation strength of at least $75\%$. We can conclude that the rotation amplitudes observed in present-day GCs are a small fraction of their primordial rotation and Eq. \ref{eq:rot_ev} offers a means to quantify it.

\subsection{Size and density of GCs at high redshift}
Figure \ref{fig:highz} shows the evolutionary tracks of our models in the surface density versus half-mass radius plane compared with current observations of real GCs. Specifically, we plot MW GCs (from the Baumgardt Galactic GCs Database\footnote{The surface densities are calculated using half of the total mass and the half-mass radius provided in the catalog, \url{https://people.smp.uq.edu.au/HolgerBaumgardt/globular/}}) and high-z observations of proto-GCs (Cosmic Gems, \citealp{Adamo2024}; Sunrise, \citealp{Vanzella2023}; Firefly Sparkle, \citealp{Mowla2024}). The evolutionary tracks of our simulations illustrate the effect of long-term evolution on GC structure: after 12 Gyr of evolution, the clusters have significantly expanded and reduced their surface density. Our models occupy the same parameter space as real GCs, suggesting that the observed high-z clusters could follow similar evolutionary tracks.

The right plot of Fig. \ref{fig:highz} shows a normalized version of the left panel: despite the different initial conditions and tidal histories of each simulation, the evolutionary tracks are very similar. After 12 Gyr of evolution, the surface density of a GC has decreased by approximately 2 orders of magnitude (a factor of $\approx1-6\times10^{-2}$), while the half-mass radius has increased by a factor of $\approx3-5$. GCs with a higher initial filling factor have lost significantly more mass (see Fig. \ref{fig:massloss_colorcoded}), and therefore they tend to have lower surface densities at a given half-mass radius. At 12 Gyr our models span a range in surface densities and sizes, which is the result of the specific initial condition and tidal history of each model. The only simulation experiencing a decrease in half-mass radius, at an approximately constant surface density, is the model \texttt{250k-A-R2-5}, characterized by strong mass loss (as discussed in Sect. \ref{sec:long_term}).

It is important to note that all our simulations are evolved in a simplified tidal field, specifically a circular orbit around a point-mass galaxy. More complex orbits within a realistic galactic potential would lead to a mass-loss rate that strongly depends on the orbit and local density. This would produce evolutionary tracks in the surface density versus half-mass radius plane that are nonlinear and could significantly enhance mass loss, particularly in the presence of strong tidal shocks—for example, those induced by disk crossings. Additional uncertainties may be due to the peculiar early evolution of high-z GCs due to the complexity of the time evolving tidal field (e.g., environment in the vicinity of their clumpy formation, buildup of host galaxy). Furthermore, all our simulations are initialized with half-mass radii of $1-4$ pc, and are designed to reproduce low-density MW GCs. Higher initial densities could also affect the evolutionary tracks by allowing clusters to remain more compact throughout their evolution. Nevertheless, our comparison observations-simulations represents a first step toward understanding the evolution of GCs from high redshift to the present day, using realistic number of particles and incorporating all relevant dynamical and stellar evolutionary processes. Future comparisons with new data (e.g., \citealp{Claeyssens2026}) will provide a critical benchmark for validating and improving theoretical models of GC evolution.

%-------------------------------------- 
\section{Initial comparison with kinematic data}
\begin{figure*}
   \centering
   \includegraphics[width=1\textwidth]{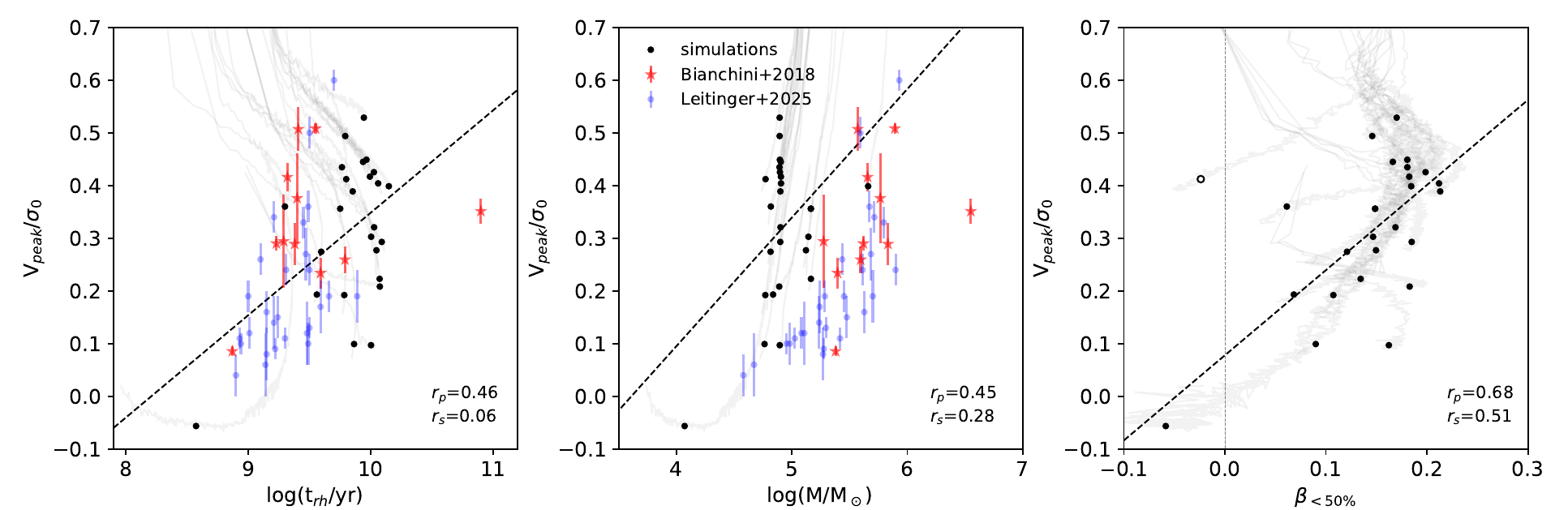}
   \caption{Comparison of our simulations (at 12 Gyr, black circles) with recent observations (\citealp{Bianchini2018b,Leitinger2025}). From left to right: $V_{peak}/\sigma_0$ vs. relaxation time at the half-mass radius, total mass, and anisotropy within the half-mass radius. The dotted lines are the linear fits to the simulation data, for which the Pearson and Spearman correlation coefficient are indicated in the bottom right. The strongest correlation was obtained for $V_{peak}/\sigma_0$ vs. $\beta_{<50\%}$, for which data are not available. In this panel, we also note that the simulation \texttt{250k-A-R4-10-retr}, characterized by retrograde rotation and strong tidal field, is a clear outlier (empty circle).}
    \label{fig:observations}%
    \end{figure*}

%-------------------------------------- 
\label{sec:5}
In this last section we carry out a first comparison with kinematic data of MW GCs. Figure \ref{fig:observations} shows the $V_{peak}/\sigma_0$ parameter (see definition in Sect. \ref{sec:gen_propr}) as a function of the half-mass relaxation time, the total mass and the velocity anisotropy. For each of our simulations we plot the snapshot at 12 Gyr with black circles. We compare our data with observations of the $V/\sigma$ parameter of MW GCs from \cite{Bianchini2018b} and \cite{Leitinger2025}. The $V/\sigma$ parameter in \cite{Leitinger2025} is estimated using the global rotation amplitude in 3D and the associated velocity dispersion both in the plane-of-the sky and along the line-of-sight. \cite{Bianchini2018b} uses the same definition as in our current work, the peak of the rotation profiles divided by the central velocity dispersion, but their values refer only to proper motions in a reference system projected on the plane of the sky. Therefore, it should be emphasized that projection effects and differences in the kinematic definitions may introduce discrepancies between models and observations. It should also be remembered that our simulations are restricted to the parameter space of lower-density MW GCs, and therefore they do not encompass the full range of observed relaxation times and masses. For this reason, we regard the present comparison with the data as an initial step.

Nevertheless, the simulations and the data exhibit broadly consistent trends, with more massive clusters (i.e., clusters with longer relaxation time) showing higher $V/\sigma$. In the case of velocity anisotropy, intrinsic estimates of the $\beta_{<50\%}$ parameter are not currently available and estimates in the literature typically only refer to the ratio of the projected tangential and radial components of proper motion velocity dispersion. These anisotropy estimates depend in a nontrivial manner on projection effects and on the inclination angle of the rotation axis. We therefore refrain from comparing the anisotropy with observations, as a meaningful assessment would require projecting our simulations into the observational plane, which lies beyond the scope of this study. However, we note that our simulations predict a clear relation between $V/\sigma$ and anisotropy. This correlation yields Pearson and Spearman coefficients of $r_p=0.68$ and $r_s=0.51$, respectively, corresponding to a strong linear association and a moderate monotonic trend.
The only noticeable outlier of this correlation (empty circle) is the retrograde model \texttt{250k-A-R4-10-retr} which is characterized by retrograde rotation coupled with a strong tidal field. An in depth analysis of this rotation versus anisotropy correlation offers a valuable testbed for future observations and motivates further efforts toward a comprehensive 3D kinematic analyses of GCs. 

%-------------------------------------- Two column figure (place early!)
%

%-------------------------------------- Two column figure (place early!)

\section{Conclusions}
\label{sec:6}
We have introduced a new suite of 25 direct N-body simulations of rotating GCs, the \texttt{ROLLIN'} simulations, specifically designed to explore the impact of internal rotation in GCs for a wide range of initial conditions. Our models are characterized by a realistic number of particles (250k-1.5M stars) and are modeled on a star-by-star basis, including stellar evolution and interaction with the external environment, using the direct summation code \texttt{NBODY6++GPU}. With this method, all relevant evolutionary ingredients, inducing collisional dynamics, are taken into consideration. 
In this paper, we investigated the evolution of the kinematic properties of our models. Our analysis has demonstrated the following:
\begin{enumerate}
    \item The early evolution of GCs is affected by the presence of internal rotation. Stronger rotating GCs undergo earlier and deeper core collapse in the first hundred million years of evolution, a finding consistent with the recent work of \cite{Kamlah2022b}. This evolutionary phase is particularly efficient in segregating massive stars and stellar remnants in the center of clusters. Fast rotating GCs reach a higher level of mass segregation early in their evolution. This also suggests the importance of rotation in setting the process of segregation of BHs in a GC center. Further studies to disentangle the combined effects of rotation and GC structural properties (e.g., their concentration) are needed to fully understand the impact on the retention of compact objects. Such investigations could provide key constraints for interpreting observations of gravitational wave sources originating from dense stellar environments. 
    \item Mass loss, originating from both stellar evolution and tidal stripping, is the main driver of the evolution of kinematic properties. While the clusters evolve, their angular momentum dissipates, and the rotation strength, quantified as the peak of the rotation profile, decreases. Independently of the initial conditions and evolution, all clusters show a rotation peak around the half-mass radius, $r_{50\%}$, as well as shapes and amplitudes of the rotation profiles that are fully consistent with observations of today's MW GCs. After 12 Gy of evolution a correlation between cluster rotation and cluster mass builds up, with more massive clusters retaining higher rotation signatures than less massive clusters, consistent with rotation measurements of MW GCs (e.g., \citealp{Bianchini2018b,Leitinger2025}).
    \item The tidal history of a GC is another key factor in the evolution of rotation profiles. Clusters with prograde rotation (with respect to their orbital angular momentum) reduce their rotation strength at a quicker pace than clusters with retrograde rotation. This is due to the preferential stripping of stars in prograde orbits, which are energetically less stable. Prograde GCs are characterized by a $V/\sigma$ profile decreasing toward the outer regions with signatures of counter rotation. Retrograde clusters are instead characterized by a more extended and pronounced rotation profile and a flat (or increasing) $V/\sigma$ profile. The differences between retrograde and prograde GCs are already visible well within the Jacobi radius. The shape of present-day rotation profiles may therefore contain important insights into the orbital history of a GC.
    \item The evolution of velocity anisotropy is driven by mass loss. All clusters are initially close to isotropy and develop radial anisotropy, peaking around 40\% mass loss (from both stellar escapers and stellar evolution), before progressing toward isotropy or tangential anisotropy for increasing mass loss. The GCs that enter the tangential anisotropy regime have undergone mass loss of $\sim60-80\%$. These systems are characterized by large initial filling factors ($r_{50\%}/r_j\approx0.05$, in the underfilling regime), making them more sensitive to the external tidal field and resulting in higher mass loss rates. The combined signature of tangential anisotropy and the lack of internal rotation can be indicative of a cluster approaching disruption.
    \item The present-day rotation of GCs is at least a factor of five lower than the rotation at formation. From the evolutionary tracks of our simulations, we derived an empirical relation (eq. \ref{eq:rot_ev}) that allowed us to reconstruct a GC's primordial rotation based on its present rotation strength and its mass loss. As an example, a mass loss of 40\% (i.e., the typical value of mass loss due to stellar evolution alone throughout a cluster's life) implies a decrease of the rotation peak of $\approx75\%$. This indicates that internal rotation must have been a key dynamical ingredient already at the earliest phases of the life of a GC. 
    \item Our simulations provide a way to compare the structural properties of old MW GCs with those of proto-GCs observed at high-z. The evolutionary tracks of our models reveal that the combined effects of stellar and dynamical evolution cause clusters to expand by a factor of about three to five. This expansion is coupled with a decrease in surface density of up to two orders of magnitude (a factor of $\approx1-6\times10^{-2}$), which is additionally driven by mass loss. These findings suggest that the observed high-redshift GCs could evolve into clusters with properties resembling those seen in the MW today. 
\end{enumerate}

In summary, we have shown that internal rotation is a key ingredient in the evolution of GCs, playing a fundamental role even at the earliest stages of their formation. Reconstructing the role of rotation in the life of a GC is therefore essential for gaining insights into GC origin, which remains largely unknown, even in the context of formation and kinematic evolution of multiple stellar populations. 

The qualitative agreement between the \texttt{ROLLIN'} simulations and the available kinematic observations of GCs indicates that these models provide a valuable benchmark for interpreting present-day as well as future observations. However, a thorough comparison between models and observations will require either investigating the projected properties of our simulations or performing a comprehensive 3D analysis of the kinematics of MW GCs—which is now possible thanks to the combination of large astrometric and spectroscopic surveys.

Finally, while our simulations mark a significant advance toward state-of-the-art direct models with a realistic number of stars and broad parameter coverage, a few key improvements remain to be made. In particular these involve (i) the inclusion of primordial binaries in the initial conditions, which is the current computational bottleneck; (ii) the adoption of more realistic time-dependent tidal fields to place GCs within a proper galactic and cosmological context; (iii) the use of less idealized initial conditions capable of capturing the complexity of cluster formation arising from the intricate interplay between gas and stars; and (iv) the inclusion of multiple stellar populations known to play a primary role already in the earliest stages of GC formation. Significant progress in these areas is already occurring (e.g., \citealp{ArcaSedda2024, Cournoyer-Cloutier2024, KaramSills2024, Lacchin2024, Webb2024, Lahen2025, Lahen2025b, Aros2025, Giersz2025, Berczik2025}), thus paving the way for a deeper understanding of the origin of GCs and their relation to high-redshift star formation and to galactic evolution.

\section*{Data availability}
The simulation outputs from the \texttt{ROLLIN'} suite are available from the corresponding author upon reasonable request. 

\begin{acknowledgements}
The set of simulations was run for a total of $\approx350\,000$~GPU hours. This corresponds to a carbon footprint of~$\approx9$~tons of CO$_{2,\mathrm{eq}}$, estimated using the methodology developed by the Labos~1point5 collective\footnote{\url{https://labos1point5.org/les-rapports/estimation-empreinte-calcul}}.

The authors thank the referee for their very useful insights. PB gratefully acknowledges Hugo Spitz and Gabriel Bounias for their valuable contributions to the visualization of simulations during their master’s internships. PB would also like to thank Rodrigo Ibata and Christian Boily for insightful discussions, and Jonathan Chardin for contributing to the early development of this work. This project was provided with computing resources by GENCI at IDRIS thanks to the following time allocations on the supercomputer Jean Zay (V100 partition): Grand Challenge-101470, A10-A0100412451, and A13-A0130412451. The authors would also like to acknowledge the High Performance Computing Center of the University of Strasbourg for supporting this work by providing access to computing resources. Part of the computing resources were funded by the Equipex Equip@Meso project (Programme Investissements d'Avenir) and the CPER Alsacalcul/Big Data. PB and AM acknowledge financial support by the IdEx framework of the University of Strasbourg. ALV acknowledges support from a UKRI Future Leaders Fellowship (MR/S018859/1; MR/X011097/1).
AA acknowledges support for this paper from project No. 2021/43/P/ST9/03167 co-funded by the Polish National Science Center (NCN) and the European Union Framework Programme for Research and Innovation Horizon 2020 under the Marie Skłodowska-Curie grant agreement No. 945339. This research was also funded in part by NCN grant number 2024/55/D/ST9/02585. For the purpose of Open Access, the authors have applied for a CC-BY public copyright license to any author Accepted Manuscript (AAM) version arising from this submission.
\end{acknowledgements}

\bibliographystyle{aa}
\bibliography{biblio} 

\begin{appendix}
\onecolumn
\section{Evolution of Lagrangian radii and mean enclosed mass}
\label{appA}
\begin{figure*}[ht!] 
   \centering
   \includegraphics[width=0.95\textwidth]{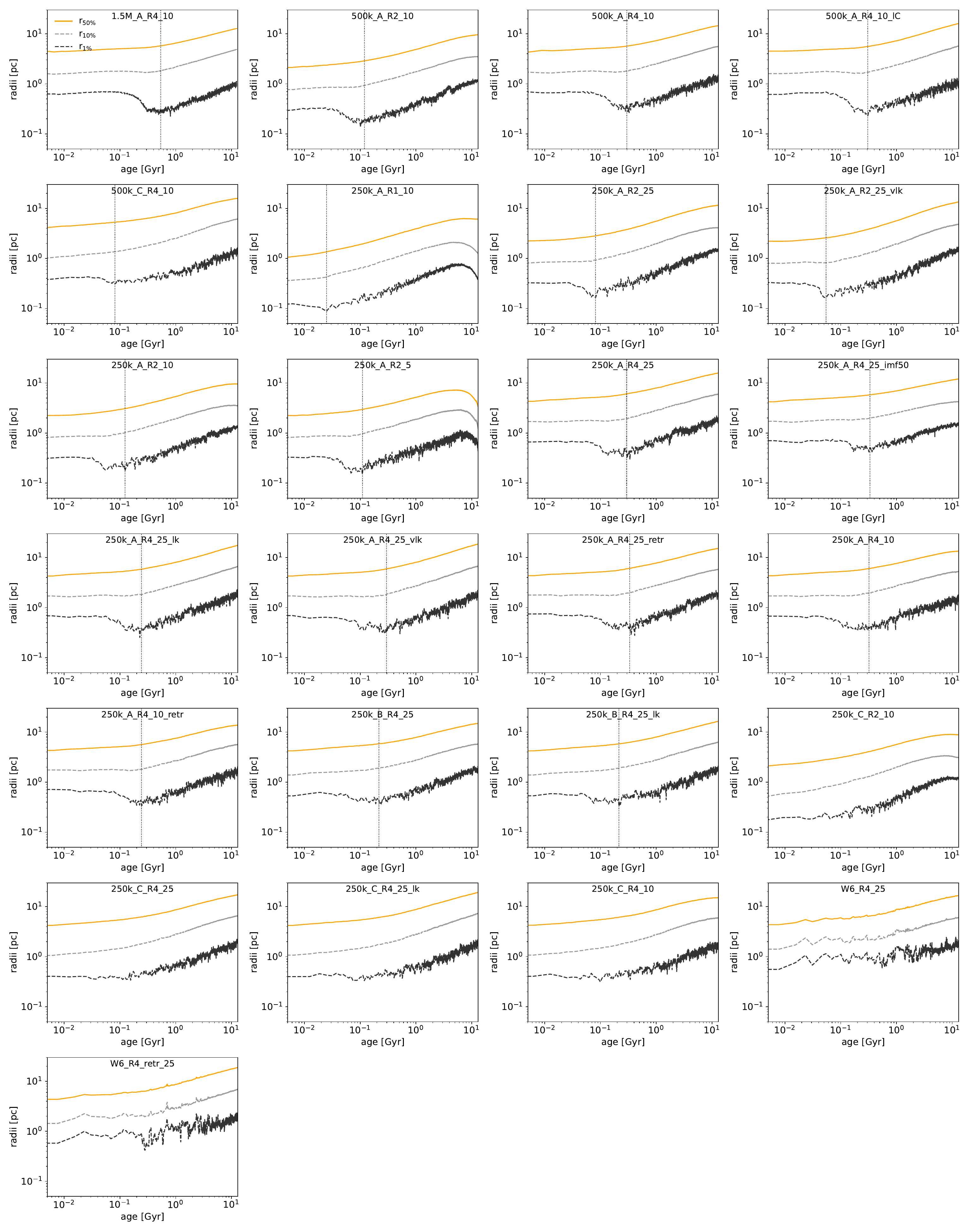}
   \caption{Evolution of 50\%, 10\% and 1\% Lagrangian radii (orange, gray, and black lines respectively) for all models, as in Fig. \ref{core_collapse}. The time of core collapse is indicated with a vertical line; when the core collapse is not clearly definable, no line is plotted.}
              \label{fig:cc_rad}%
    \end{figure*}

\begin{figure*}[ht!] 
   \centering
   \includegraphics[width=0.95\textwidth]{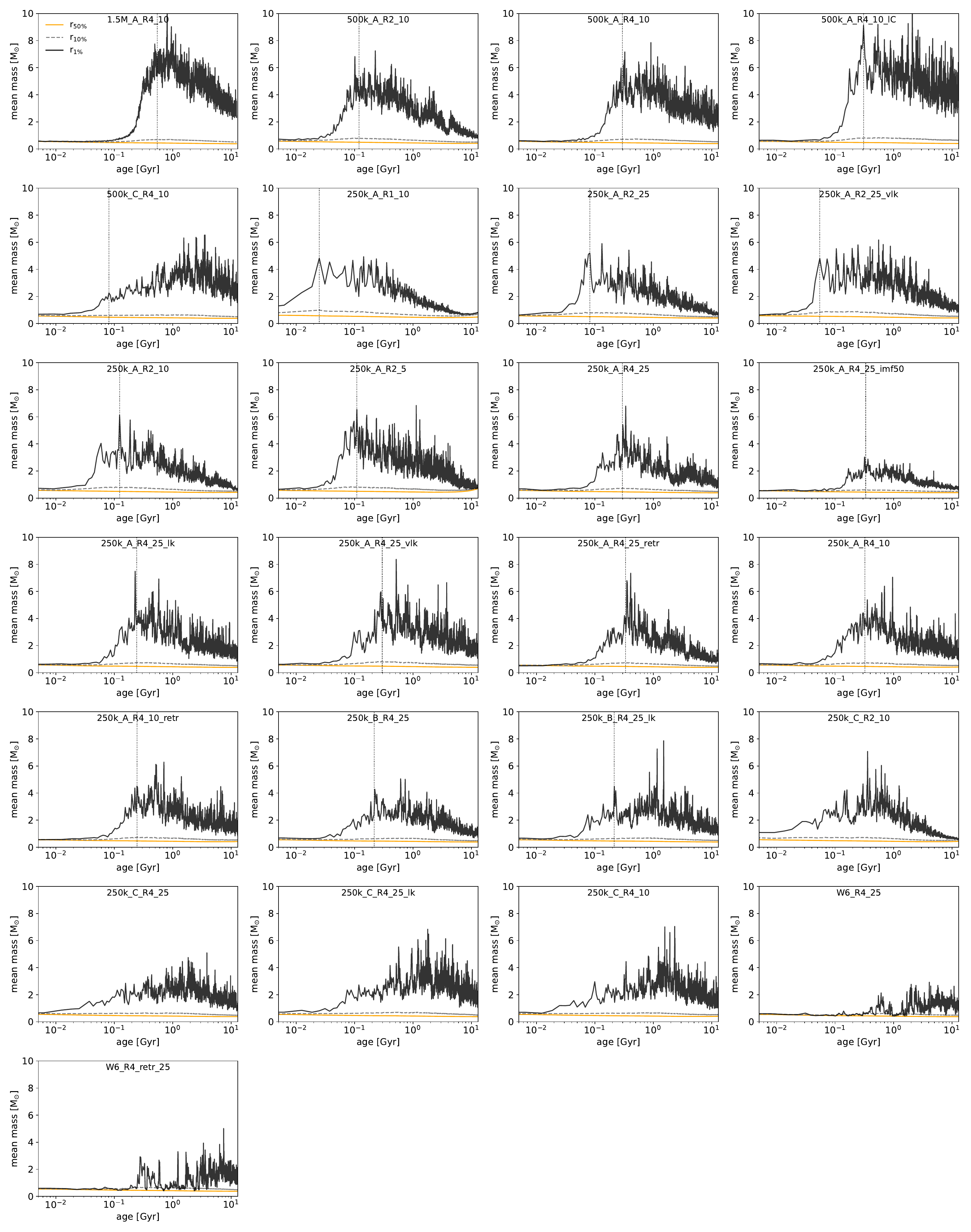}
   \caption{Evolution of the mean mass within the 50\%, 10\% and 1\% Lagrangian radii (orange, gray, and black lines respectively) for all models, as in Fig. \ref{core_collapse}. The time of core collapse is indicated with a vertical line; when the core collapse is not clearly definable, no line is plotted.}
              \label{fig:cc_mass}%
    \end{figure*}

Evolution of Lagrangian radii (Fig. \ref{fig:cc_rad}) and mean enclosed mass (Fig. \ref{fig:cc_mass}) for all the models in our suite of simulations.

%-------------------------------------- Two column figure (place early!)
%

%-------------------------------------- Two column figure (place early!)
%-------------------------------------- 

\end{appendix}

\end{document}